\documentclass[12pt,preprint]{aastex}

\usepackage{epsf}

\shorttitle{Galaxy Winds and the Ly$\alpha$ forest}
\shortauthors{Kollmeier et al.}
\newcommand{\etal}{et al.\ }

\newcommand{\lya}{Ly$\alpha$\ }
\newcommand{\kms}{\,{\rm km}\;{\rm s}^{-1}}

\newcommand{\hmpc}{\,h^{-1}\;{\rm Mpc}}

\newcommand{\msun}{M_\odot}
\newcommand{\zsun}{Z_\odot}
\newcommand{\lsun}{L_\odot}

\newcommand{\Mpc}{{\rm Mpc}}

\newcommand{\be}{\begin{equation}}
\newcommand{\ee}{\end{equation}}

\begin{document}
\title{Galactic Wind Effects on the \lya Absorption in the Vicinity of Galaxies}
\author{Juna A. Kollmeier$^1$ Jordi Miralda-Escud\'e$^{1}$, Renyue Cen$^2$, \& Jeremiah P. Ostriker$^2$}

\footnotetext[1]
{Ohio State University, Dept.\ of Astronomy, Columbus, OH 43210,
jak,jordi@astronomy.ohio-state.edu}
\footnotetext[2]
{Princeton University, Dept.\ of Astronomy, Princeton NJ,
cen, jpo@princeton.edu}

\begin{abstract}
  We present predictions of \lya forest-galaxy correlations at $z=3$
from Eulerian simulations that include the effects of galactic winds,
driven primarily by supernova explosions. Galactic winds produce
expanding bubbles of shock-heated gas within $\sim 0.5$ comoving
$\hmpc$ of luminous galaxies in the simulation, which have space
density similar to that of observed Lyman break galaxies
(LBGs). However, most of the low-density intergalactic gas that
determines the observed properties of the \lya forest is unaffected by
winds. The impact of winds on the \lya optical depth near galaxies is
less dramatic than their impact on gas temperature because winds heat
only a small fraction of the gas present in the turnaround regions
surrounding galaxies. Hence, \lya absorption from gas outside the wind
bubbles is spread out over the same velocity range occupied by the
wind-heated gas. In general, \lya absorption is expected to be
stronger than average near galaxies because of the high gas
density. Winds result in a modest reduction of this expected increase
of \lya absorption. Our predictions can be compared to future
observations to detect the wind effects and infer their strength,
although with the caveat that the results are still dependent on the
correspondence of simulated galaxies and observed LBGs.  We find that
wind effects in our simulations are not strong enough to reproduce the
high \lya transmission within 0.5 $\hmpc$ comoving of galaxies that
has been suggested by recent observations; powerful galactic
explosions or ejecta with hyper-escape velocities would be required,
but these are unlikely to be produced by ordinary star formation and
supernovae alone.

\end{abstract}
\section{Introduction}
\label{sec:intro}

  Galactic winds are an important but poorly understood aspect of
galaxy formation and evolution. Winds are observed locally (e.g.,
\citealt{heckman98, martin98}), but are probably more important at
high redshift where star formation and the subsequent supernovae occur
at the fastest rate. Winds have been invoked as an important force for
forming galaxies \citep{ostrikerandcowie81}, and more recently their
effects have been proposed as a solution to various problems related
to galaxy formation, from the suppression of star formation in low
mass halos (now the dark dwarf satellites predicted to orbit galaxies
like the Milky Way; e.g., \citealt{dekel86}), to the metal pollution
of the intergalactic medium (hereafter, IGM; see, e.g.,
\citealt{aguirre01}). Luminous, star-forming galaxies at high
redshift, or Lyman break galaxies (hereafter, LBGs) show evidence of
outflows in their spectra \citep{pettini02} with velocities of up to
1000 $\kms$.

  Such outflows may not only affect star formation within the galaxy,
but may potentially impact the surrounding intergalactic gas as well.
Recently, \citet{adelberger03} have reported a ``galaxy proximity
effect'' in the IGM around LBGs. The effect consists
of an increase in the \lya transmitted flux within a comoving
transverse separation of $\sim 1 \hmpc$ and at the redshift of the LBG
as measured on a parallel line of sight probed by a quasar. This is
contrary to the expected {\it decrease} of transmitted flux (or
increased absorption) owing to increased gas density in the vicinity
of a galaxy predicted by the gravitational instability model, which
has been observed at low redshift (e.g., Morris \etal 1993; Lanzetta
\etal 1995; Chen \etal 1998). \citet{adelberger03} interpreted their
observations as indicating the effects of galactic super-winds that
would have either swept up the absorbing gas, or otherwise heated it
sufficiently to remove its absorption signature entirely.

  The strength of a galactic wind required to remove all of the
hydrogen \lya absorption from the region surrounding a galaxy is,
however, more extreme than one might at first guess. The absorbing
hydrogen must be removed at least out to the turnaround radius of the
galactic halo because the peculiar velocity of any gas at that
distance will bring its \lya absorption line to the same redshift as
the galaxy, and the gas density at the turnaround radius is typically
already considerably larger than the mean density of the universe,
enough to cause strong absorption. In a galactic halo with velocity
dispersion $\sigma$, the material that has formed the galaxy must
originally have collapsed by falling from the turnaround radius with a
velocity $\sim \sigma$ during the age of the universe, $t$. Hence, a
wind of lifetime $t_w$ must move out at a velocity $\sim \sigma t/t_w$
to reach out to the same radius and sweep all the gas away. An exact
calculation for a spherical top-hat perturbation in which the halo
velocity dispersion remains constant with time yields a minimum wind
velocity of $(\sqrt{8}\sigma/\pi)t/t_w$ to reach the turnaround
radius, or even larger if the halo velocity dispersion increases with
time, as is the case for most halos in the cold dark matter (CDM)
model. In practice, the wind lifetime should be a small fraction of
the age of the universe because the wind removes the gas falling onto
the halo from large distances that is needed to sustain a long period
of star formation. This implies that the wind must have moved at a
``hyper-escape'' velocity: the wind must reach the turnaround radius
(already the intergalactic region surrounding the halo) and still be
moving at a speed larger than the halo velocity dispersion. In other
words, the expelled gases would actually have to be the debris from an
explosion, rather than a self-regulated wind in which the gas is
expelled at the speed required to escape the potential well of the
halo.

  It appears unlikely that this type of explosion would generally have
occurred in all high-redshift LBGs, both because we would expect winds
from star-forming galaxies to be self-regulated (in other words, the
star formation rate should decrease as a wind is able to expel the
infalling gas that feeds the star formation) and because of the
enormous energy required to remove the absorbing gas out to more than
$0.5 \hmpc$ from LBGs \citep{adelberger03}. Because the observational
result is based on only a few galaxies, one must be cautious with this
interpretation.

  Numerical simulations have shown that an increased transmitted flux
near galaxies, at the level observed by \citet{adelberger03}, is only
produced by extreme feedback models. A thermal feedback effect is too
weak to sufficiently reduce the absorption of the dense gas near a
galaxy \citep{croft02, kollmeier03a}. Simple wind models in which LBGs
were associated with massive galaxies at high redshift \citep{croft02,
kollmeier03a, kollmeier03b, desjacques04} have shown that gas must be
affected at unrealistically large distances from galaxies to result in
increased \lya transmission. More sophisticated wind model
prescriptions \citep{croft02, bruscoli03} do not match the
observations of \citet{adelberger03} either. Photoionization by a
massive galaxy itself does not produce an intensity of ionizing
radiation that is large compared to the cosmic background
\citep{croft02, kollmeier03a}. 

  In this paper we investigate the effect of galactic winds on model
predictions for the statistics of the transmitted flux of the \lya
forest close to the galaxies from which the winds originate. To date,
these predictions have not been made using Eulerian cosmological
simulations, although \cite{mcdonald02} do check the robustness of
their HPM runs using simulations of the type presented here. We
explore the effect of winds using this type of simulation
\citep{cen04}, which treat winds very differently from previous
studies. Furthermore, we are able to probe the wind effects directly
since we compare three different simulations in which only the wind
strength is varied.  We also make predictions for new statistical
measures of the \lya absorption near galaxies that may help better
compare observations and theory.  In \S 2 we describe the simulations,
including their star formation and wind prescriptions and our methods
of identifying galaxies in the simulation and of analyzing the \lya
forest near them.  In \S 3 we present a picture of the physical
conditions of the IGM in the presence of feedback.  In \S 4 we
quantify this picture and present our main results for the flux
statistics from our simulations, and we compare our predictions with
recent observations. We present our conclusions and discuss our
results in \S 5.

\section{Simulations}
\label{sec:sims}

We analyze the outputs at $z=3$ of four Eulerian simulations of a
$\Lambda$-CDM universe.  These simulations are all improved versions
of the original simulations by \cite{cen92,cen99a,cen00} described in
detail in Cen \etal (2004) and in the other papers by this group.  In
order to understand the effect of galactic winds on our results, we
analyze three simulations of a LCDM cosmology that differ in the
amount of energy that is assumed to be released by supernovae. The
cosmological parameters of all three simulations are
$(\Omega_m,\Omega_\Lambda, \Omega_b,\sigma_8,h)=(0.29,0.71,0.047,0.85,
0.70)$, each of size $11\hmpc$ comoving on a side with $432^3$ cells. The
different amounts of input supernova energy allow us to test directly
the effect of winds produced by supernovae on our predictions for
galaxy-\lya forest correlations. We distinguish the three simulation
boxes by the designations ``L11-Low'', ``L11-Medium'', and
``L11-High'' where these correspond to input supernova efficiencies
(see below, eq.\ \ref{esn}) of $\epsilon_{SN} = 3\times 10^{-7}$,
$3\times 10^{-6}$, and $1.5\times 10^{-5}$, respectively. In addition,
we analyze a large box of size $25\hmpc$ on a side with $768^3$ cells
and feedback parameter $\epsilon_{SN}=1\times 10^{-5}$. This large box
has similar resolution as the other three simulations, but the finite
box size effects on the galaxy-\lya forest correlations should be less
severe than in the other three simulations, allowing for a more direct
comparison to the observations of Adelberger \etal (2003). We refer to
this box as ``L25''. A summary of the simulations and parameters is
given in Table 1.

\subsection{Star Formation and Supernova Energy Prescriptions}

Baryonic mass in overdense regions of the simulation is converted to
``star particles'' upon satisfying the following criteria in any given
cell: (1) converging flow ($\nabla \cdot v < 0$), (2) rapid cooling
($t_{cool} < t_{dyn}$), and (3) Jeans instability ($m_{gas} > m_{J}$)
(Cen \etal 2004). Once these conditions are met, a star particle is
formed with mass equal to $m_\ast = c_\ast m_{gas}\Delta_t/t_\ast$,
where $m_{gas}$ is the gas mass within the cell, $\Delta_t$ is the
current timestep in the simulation, and $c_\ast$ is the star formation
efficiency taken to be 0.07, except for the L25 run in which
$c_\ast=0.25$ is used.  The choice of $c_\ast$ is made to match two
observations: the total stellar density at $z=0$ (e.g.,
\citealt{fukugita98}) and the reionization epoch at $z\sim 6$ (e.g.,
Cen \& McDonald 2002).  The value of $t_\ast$ is determined by the
local dynamical time and is equal to the larger of the dynamical time
and $10^7$ yr (Cen et al. 2004); the lower bound of $10^7$ yr is
imposed somewhat arbitrarily to reflect a possible minimum star
formation timescale in order to smooth out star formation over time,
although the results do not sensitively depend on it.  A galaxy within
the simulation is simply a group of these star particles.  We discuss
the procedure we use for galaxy identification by linking star
particles into groups in the next section. The prediction for the star
formation rate is reasonably robust to variations in the adjustable
parameter $c_\ast$ for a given flux of gas cooling into a galaxy.
Changes in $c_\ast$ may change the reservoir in gas being transformed
into stars but have little effect on the rate of star formation, which
tends to match the infall rate.

  The formation of stars and active galactic nuclei (hereafter, AGN)
will generally result in two feedback effects on the gas that can form
stars: heating by radiation and mechanical energy injection from
stellar winds and supernova explosions. The magnitude of these effects
in different environments is highly uncertain, and they are
parameterized in the simulation by the efficiency parameters
$\epsilon_{\ast}$, $\epsilon_{AGN}$, and $\epsilon_{SN}$. Stars and
AGN contribute to a background radiation field (assumed to be always
of uniform intensity throughout the simulated box) as
\begin{eqnarray}
\Delta E_{rad,\nu} = m_\ast c^2
(\epsilon_{\ast} g_{\nu} + \epsilon_{AGN} f_{\nu}) ~,
\end{eqnarray}
where $m_\ast$ is the total mass of stellar particles that are formed
in the simulation during a given timestep, and $\Delta E_{rad,\nu}$ is
the energy added to the background radiation in the box over the same
timestep per unit frequency. The functions $g_{\nu}$ and $f_{\nu}$ are
the spectral energy distributions of a young stellar population and an
AGN spectrum \citep{edelson86}, respectively (normalized so that their
integral over frequency is unity). We use the Bruzual-Charlot
population synthesis code (Bruzual \& Charlot 1993; Bruzual 2000) to
compute the intrinsic metallicity-dependent UV spectra from stars with
Salpeter IMF (with a lower and upper mass cutoff of $0.1\msun$ and
$125\msun$).  Note that $\epsilon_{\ast}$ is a function of
metallicity.  The Bruzual-Charlot code gives
$\epsilon_{\ast}=(1.2\times 10^{-4}, 9.7\times 10^{-5}, 8.2\times
10^{-5}, 7.0\times 10^{-5}, 5.6\times 10^{-5}, 3.9\times 10^{-5}
,1.6\times 10^{-6})$ at $Z/\zsun=(5.0\times 10^{-3}, 2.0\times
10^{-2}, 2.0\times 10^{-1}, 4.0\times 10^{-1}, 1.0, 2.5 ,5.0)$, and we
use an interpolation of these values in the simulation.  We also
implement a gas metallicity dependent ionizing photon escape fraction
from galaxies in the sense that higher metallicity (hence higher dust
content) galaxies are assumed to allow a lower escape fraction; we
adopt the escape fractions of $f_{esc}=2\%$ and $5\%$
\citep{hurwitz97, deharveng01, heckman01} for solar and one tenth of
solar metallicity, respectively, and interpolate and extrapolate
linearly in $f_{esc}$ and in [Fe/H]. In addition, we include the
emission from AGN using the spectral form observationally derived by
\citet{sazonov04}, with a radiative efficiency in terms of stellar
mass of $\epsilon_{AGN} = 2.5\times 10^{-5}$ for $h\nu>13.6$eV. The
ionizing radiation is released over time by each star particle
according to the following function:

\begin{equation}
 f(t,t_i,t_{dyn}) = (t-t_i)\,
 \exp \left(-{t-t_i \over t_{dyn} }\right) ~,
\end{equation}
where $t_i$ is the formation time of a stellar particle, and $t_{dyn}$
is the dynamical time of the gas in the cell in which the star particle
formed.

  For the supernovae, a similar prescription is adopted whereby the
total supernova energy released by a star particle of mass $m_\ast$ is
\begin{equation}
\Delta E_{SN} = \epsilon_{SN} m_\ast c^2 ~.
\label{esn}
\end{equation}
In contrast to the radiation energy, however, the supernova mechanical
energy and ejected matter are distributed into 27 local gas cells
centered at the stellar particle in question, weighted by the inverse
of the gas density in each cell.  The mass released back into the
medium by the supernovae is fixed to be $\epsilon_{mass} m_\ast$, with
$\epsilon_{mass}=0.25$.  The supernova efficiency factor is varied in
each of the three simulations with box size of $11 \hmpc$, with values
$\epsilon_{SN} = 3\times 10^{-7}$, $3.0\times10^{-6}$, and $1.5\times
10^{-5}$.  If the ejected mass and associated energy propagate into a
vacuum, the resulting velocity of the ejecta would be
$(2\epsilon_{SN}/\epsilon_{mass})^{1/2}c=1469 \kms$, for the
L11-Medium simulation with $\epsilon_{SN}=3\times 10^{-6}$. This
medium value of $\epsilon_{SN}$ also corresponds to releasing an
energy of $10^{51}$ ergs (roughly the energy of one supernova) for
every $200 \msun$ of stars that are formed. After the ejecta have
accumulated an additional mass comparable to their initial mass, the
velocity may slow down to a few hundred $\kms$.  We assume this velocity
would roughly correspond to the observed outflow velocities of LBGs
(e.g., Pettini \etal 2002).  The large simulation with box size of $25
\hmpc$ has a supernova energy efficiency of $\epsilon_{SN} = 1\times
10^{-5}$.  The supernova energy is not released instantaneously when a
star particle forms, but is released over time with the distribution $
d\Delta E_{SN}/dt \propto \exp [ - (t-t_i)/t_{SN} ] $, where $t_i$ is
the formation time of the star particle and $t_{SN}=10^{8}$yr is a
characteristic timescale over which supernova energy is deposited in
the IGM.

\subsection{Galaxy Identification}

\begin{figure*}
\centerline{
\epsfxsize=3.0truein
\epsfbox[28 31 270 554]{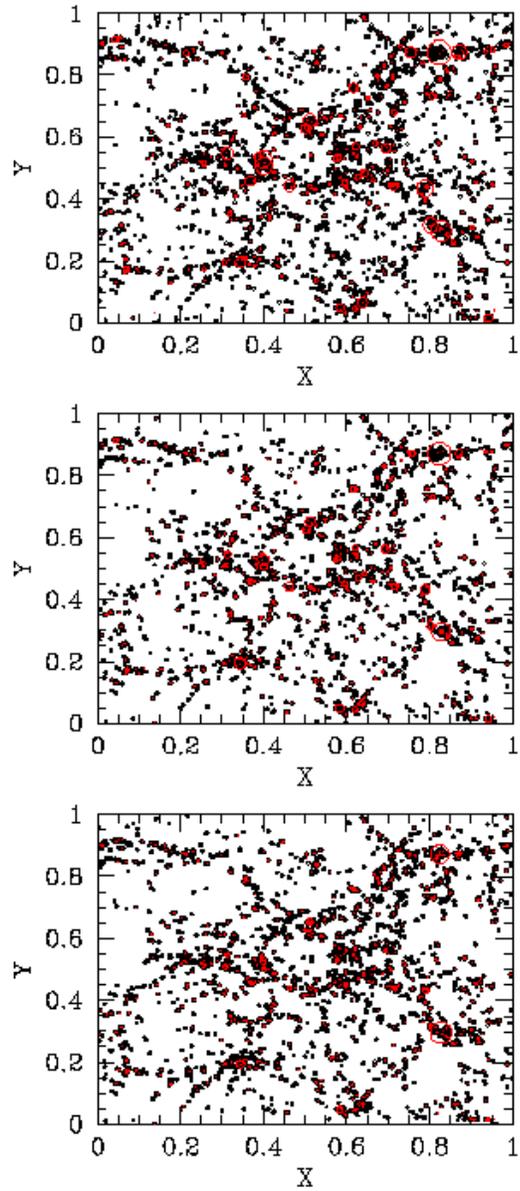}
}
\caption{Examples of our galaxy grouping method. Shown is the x-y projection in the L11-Low (top), L11-Medium (middle) and L11-High (bottom) simulations.  Black points indicate star particles and red points show galaxies with symbols whose size is proportional to stellar mass.
}
\label{fig:starparticles}
\end{figure*}

\begin{figure*}
\centerline{
\epsfxsize=6.5truein
\epsfbox[36 477 306 686]{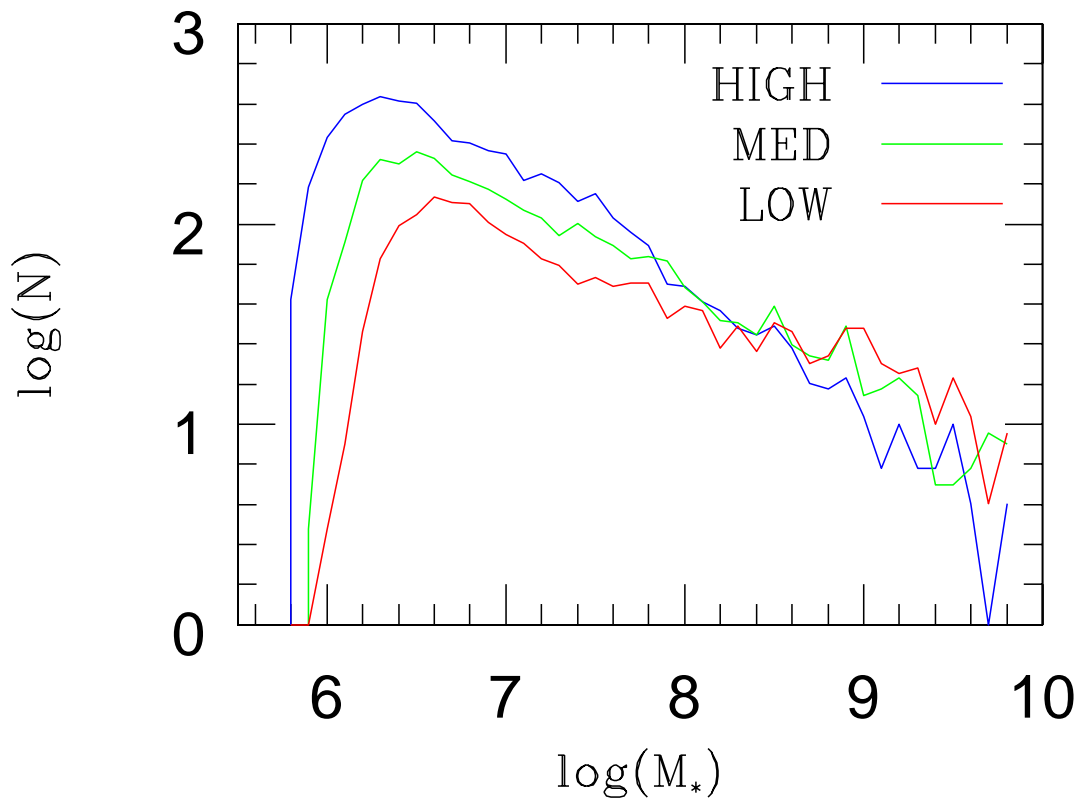}
}
\caption{ The galaxy mass function for the L11-Low(red), L11-Medium(green) and L11-High(blue) simulations. Note that L11-High has an overabundance of low mass galaxies and a deficit of high mass galaxies relative to L11-Low or L11-Medium indicating that feedback effects are ``breaking up'' galaxies.
}
\label{fig:massfcn}
\end{figure*}

  Star particles in the L11 simulations are grouped into galaxies using
a friends-of-friends algorithm with a linking length of 0.2 times the
mean interparticle spacing. We further require that each galaxy
contain at least 20 star particles. Figure \ref{fig:starparticles}
shows the positions of star particles and the resulting galaxies
projected in the x-y plane for each simulation as an illustrative
example.  We rank the galaxies in the simulation according to their
stellar mass and pick samples above thresholds in stellar mass.  There
are 1841, 2969, and 5209 galaxies in the L11-Low, L11-Medium and
L11-High simulations respectively corresponding to comoving number densities of
1.38, 2.2, and 3.9 $h^3 \Mpc^{-3}$.  The galaxy (stellar) mass function is
shown for each of the simulations in Figure \ref{fig:massfcn}. The L11-Low
simulation has more high mass objects than either the L11-Medium or L11-High
simulations, and conversely, the L11-High simulation has significantly
more low mass objects than either the L11-Low or L11-Medium simulations.
We expect the supernova energy injection to suppress star formation
owing to the heating of the gas, and indeed the total stellar mass in
the L11-Low simulation is higher than in the simulations with stronger
winds (the stellar mass is dominated by high mass galaxies).
It is unclear what process causes an increased abundance of low-mass
objects in the L11-High feedback case; the numerical implementation of
the supernova energy injection in the simulations probably tends to
break up spatial regions of star formation as more energy is injected.

  Because of the relatively small box size we use, we do not have many
rare objects (i.e., objects of low space density, as LBGs are observed
to have). We therefore cannot consider our galaxy samples for these
three simulations as LBG samples explicitly, since the estimated
number density of LBGs in a box of this size would require us to do
our calculations around only a handful of galaxies which would render
our statistics quite poor. This partly motivates our decision to
analyze additionally the larger-box L25 simulation in which we have
sufficient volume to reasonably match the observed space density of
LBGs. Star particles in this larger simulation are grouped according
to the HOP group algorithm \citep{eisenstein98} with threshold
parameters $(\delta_{outer}, \delta_{saddle},
\delta_{peak})=(80,200,240)$ and is described in detail in
\citet{nagamine01b}. We refer the reader to that paper for a more
detailed description of the galaxy identification method.

  We also use modeled B-band luminosities for the galaxies in the
simulation, in order to rank them by luminosity in \S 4 below. These
luminosities are computed by adding the emissivity of all star particles
in a galaxy, as determined from the age of the star particle and the
instantaneous burst-model isochrone-synthesis code GISSEL99
(see \citealt{nagamine00} for more details of this procedure).

\subsection{Spectral Extraction}

  For each of the boxes we analyze, we generate a grid of 1200
artificial spectra from the temperature, velocity, and density fields.
We extract 400 such sightlines along each projection of the box.  Each
spectrum has a number of ``pixels'', which we set equal to the number
of grid cells on a side of the box (432 or 768 pixels for the
simulations considered here). We compute \lya spectra from the
temperature, density, and velocity distributions of the gas in the
simulation. The spectra are normalized to a fixed mean transmitted flux
equal to the observed value of 0.638 at $z=3$ \citep{press93}. We use a
photoionization rate of $\Gamma_{\gamma,HI}=1\times 10^{-12} s^{-1}$ for
the L11 boxes and then multiply the resulting optical depths by small
factors in order to ensure that the normalization matches the observed value
of the mean flux decrement (0.36). We use the recombination
coefficient from Verner \& Ferland\ (1992), and the collisional
ionization coefficient from \cite{voronov97}. Note that collisional
ionization may be important for distinguishing the effects of winds
close to galaxies where the temperature is high.

\begin{figure*}
\centerline{
\epsfxsize=6.5truein
\epsfbox[52 188 566 700]{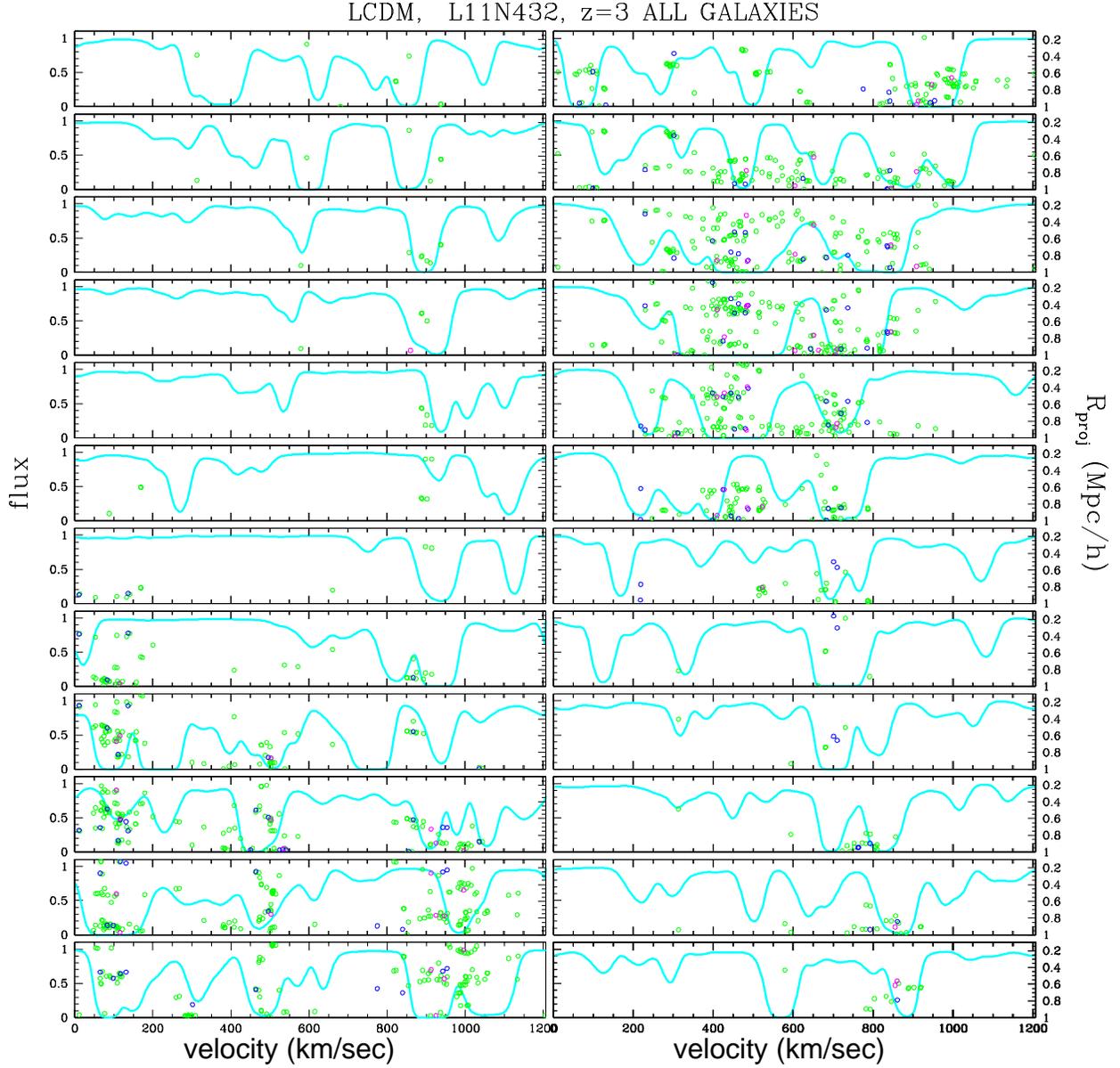}
}
\caption{Galaxy-absorption correlations in simulated spectra in the L11-Medium simulation.  Symbols show the positions of galaxies located at different redshifts along the sightline located at projected comoving distances as indicated on the right hand axis.  Green, blue, magenta symbols correspond to galaxies with masses of $10^6< M \leq 10^8, 10^8< M \leq 10^9, > 10^{9} M_\odot$ respectively.  Galaxies lying close to the line of sight are clearly associated with strong absorption features at the galaxy systemic redshift in the simulations.
}
\label{fig:losgal}
\end{figure*}

  Examples of simulated spectra from the L11-Medium simulation are shown in
Figure~\ref{fig:losgal}. The points superposed on the spectra indicate
the redshifts and projected distances (right axis of the figure) of
galaxies within $1 \hmpc$ of the sightline. The stellar mass of
each galaxy is coded by point color. We see that galaxies lying within
$0.5\hmpc$ are always associated with strong \lya absorption. Often,
clusters of galaxies are associated with large extended absorption
features spanning several hundred $\kms$. Note that strong absorption
features are not always associated with galaxies, but galaxies are
always close to some absorption feature.

\section{Physical Impact of the Winds on the IGM and the \lya Forest}

\begin{figure*}
\centerline{
\epsfxsize=6.5truein
\epsfbox[35 245 554 688]{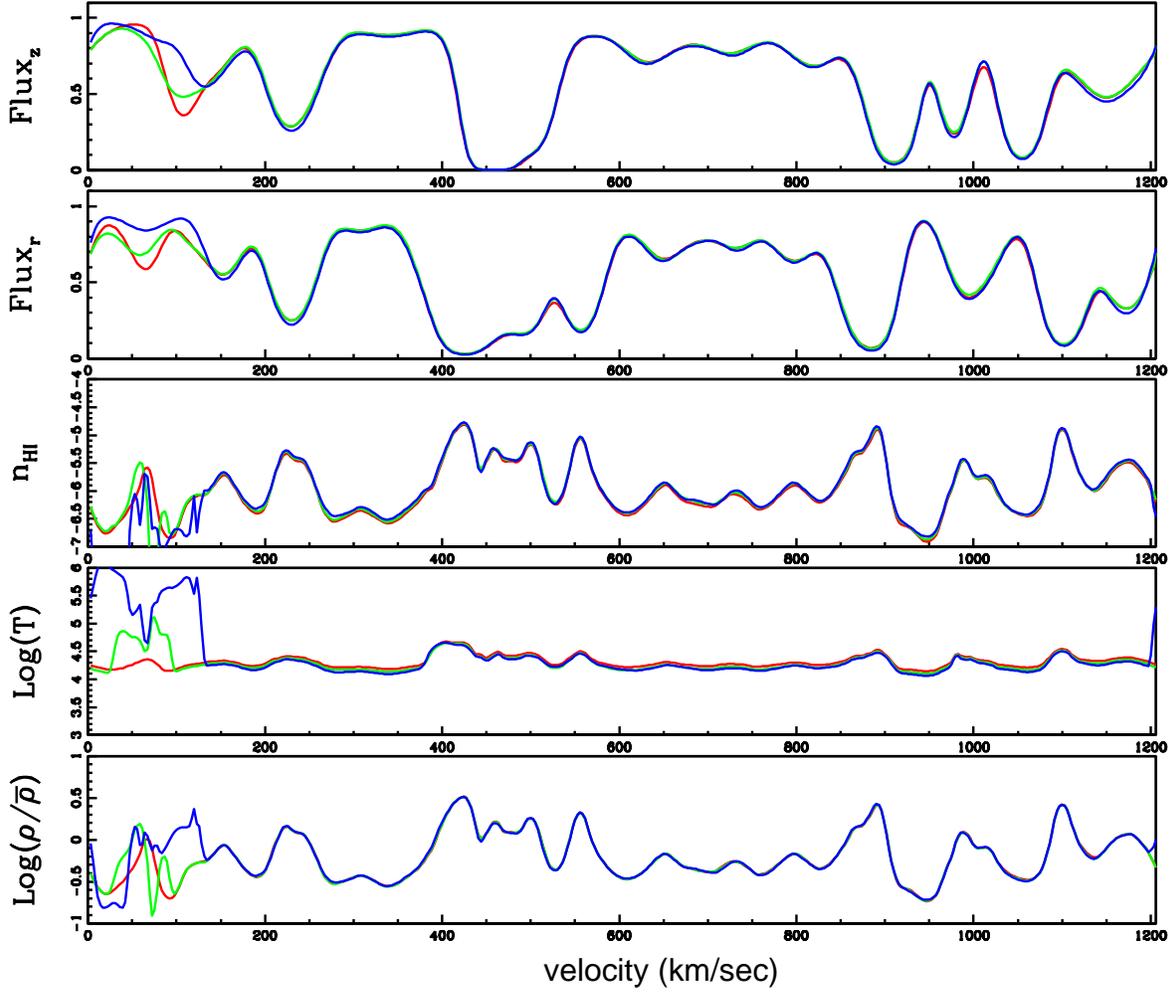}
}
\caption{Sample sightline drawn at the same location in each of the
L11-Low(red), L11-Medium(green), and L11-High(blue) simulations.
Panels (from top to bottom) show the redshift space spectrum, the real
space spectrum, the neutral fraction, temperature and density
contrast. The redshift-space (observed) flux is not dramatically
different between the three simulations despite the fact that
temperature and density differences between the sightlines are
evident.}
\label{fig:los}
\end{figure*}

  To see the typical effect of winds on the \lya\ forest spectra,
Figure~\ref{fig:los} shows spectra from a randomly chosen position in
the L11 simulations. The three curves show the same sightline in the
L11-Low(red), L11-Medium(green) and L11-High(blue) simulations. The
top panel shows the flux as observed in redshift space, and the second
panel shows the flux in real space (i.e., with gas peculiar velocities
set to zero). The third, fourth and fifth panels show the real-space
neutral hydrogen density, temperature, and gas overdensity
respectively.  The figure shows the well-known correspondence between
peaks in the gas density and absorption troughs in the \lya\ forest
spectrum. Galaxies in the simulations form in the densest central
regions of halos, and so they should clearly be
associated with deep absorption troughs up to the transverse distance
at which galactic halos are surrounded by dense gas.

  We also see in Figure~\ref{fig:los} that, over most of the spectrum,
the three simulations are almost identical, indicating that the effect
of winds is very small. The gas temperature is generally highly
correlated with gas density. However, we can clearly identify one
region of the spectrum in which the three simulations are very
different and very high temperatures are reached in regions of widely
different densities. This region has obviously been affected by
winds. The gas heated by winds can result in weak, broad \lya
features, but this effect is hard to identify and to distinguish from the
general \lya forest arising from colder gas. For example, the \lya absorption
at $v\sim 50 \kms$ in this example becomes broader and weaker with
increased wind strength, but it seems difficult to statistically
detect these changes on the \lya forest because of the low fraction of the
spectrum affected by winds.  In fact, as found by
\citet{mcdonald04}, the effects of galactic winds tend to be
degenerate with parameters affecting the typical gas temperature in
the IGM, which depends on the uncertain history of
heating during helium reionization.

\begin{figure*}
\centerline{
\epsfxsize=2.5truein
\epsfbox[14 36 377 443]{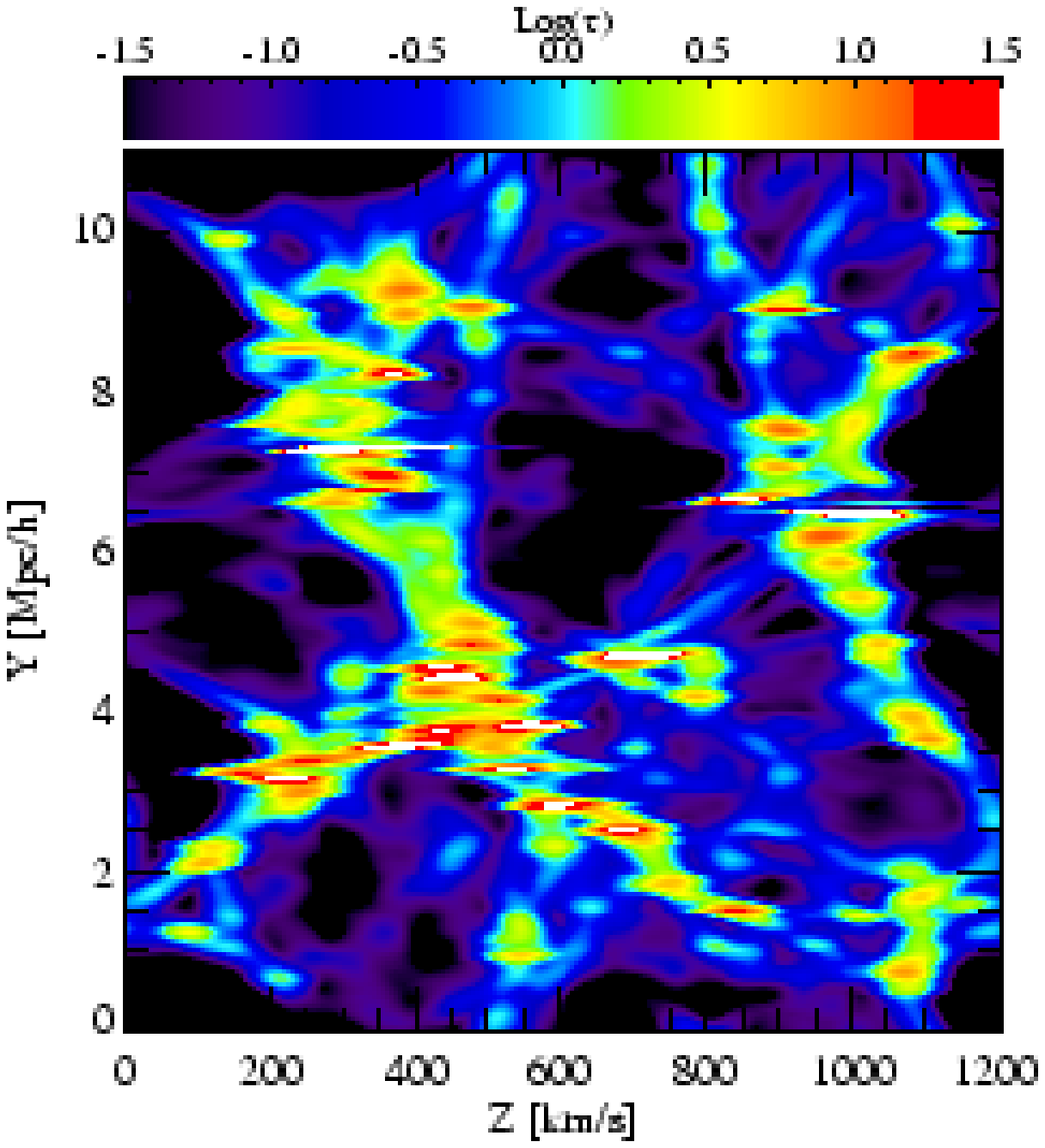}
\epsfxsize=2.5truein
\epsfbox[14 36 377 443]{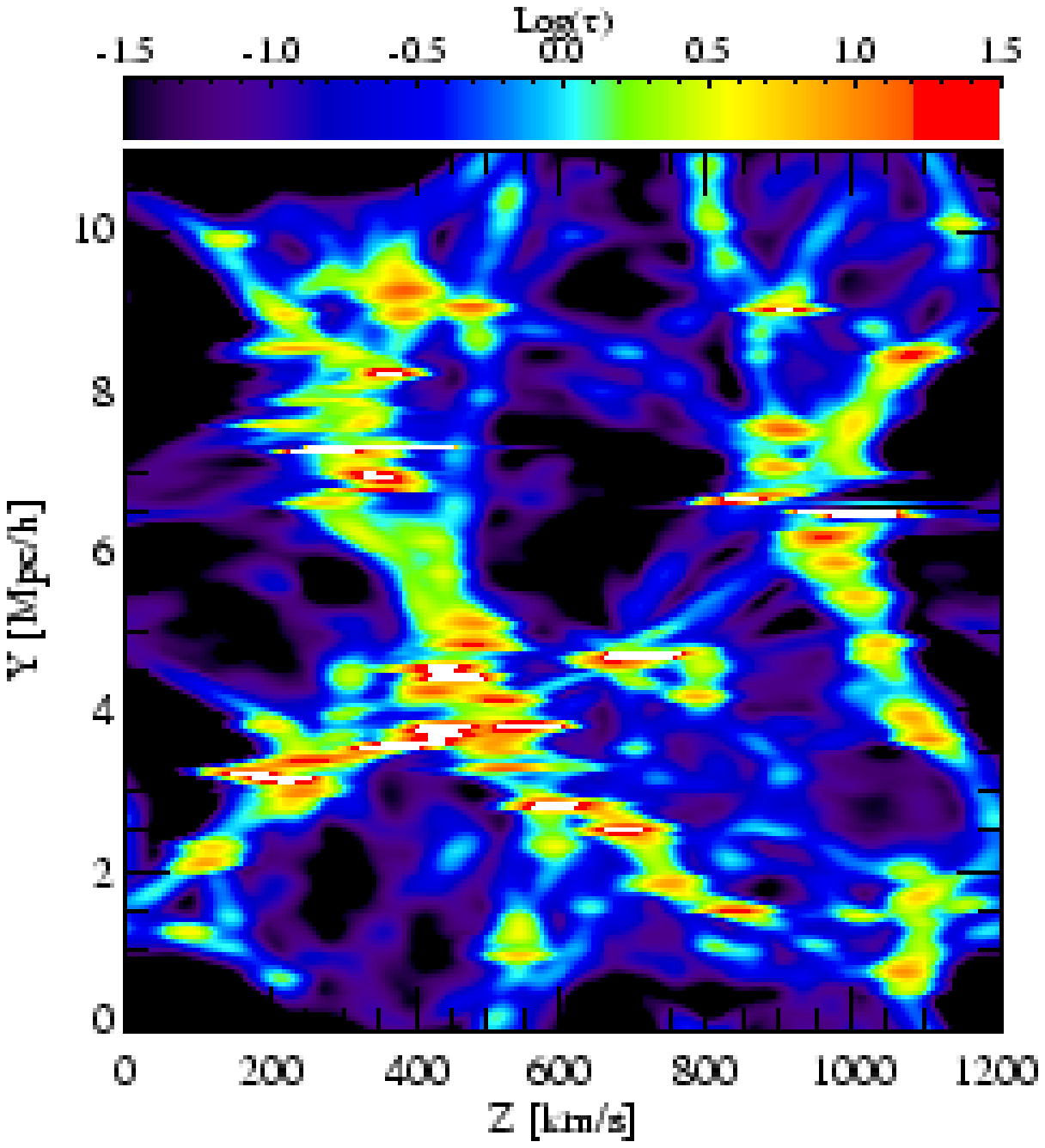}
\epsfxsize=2.5truein
\epsfbox[14 36 377 443]{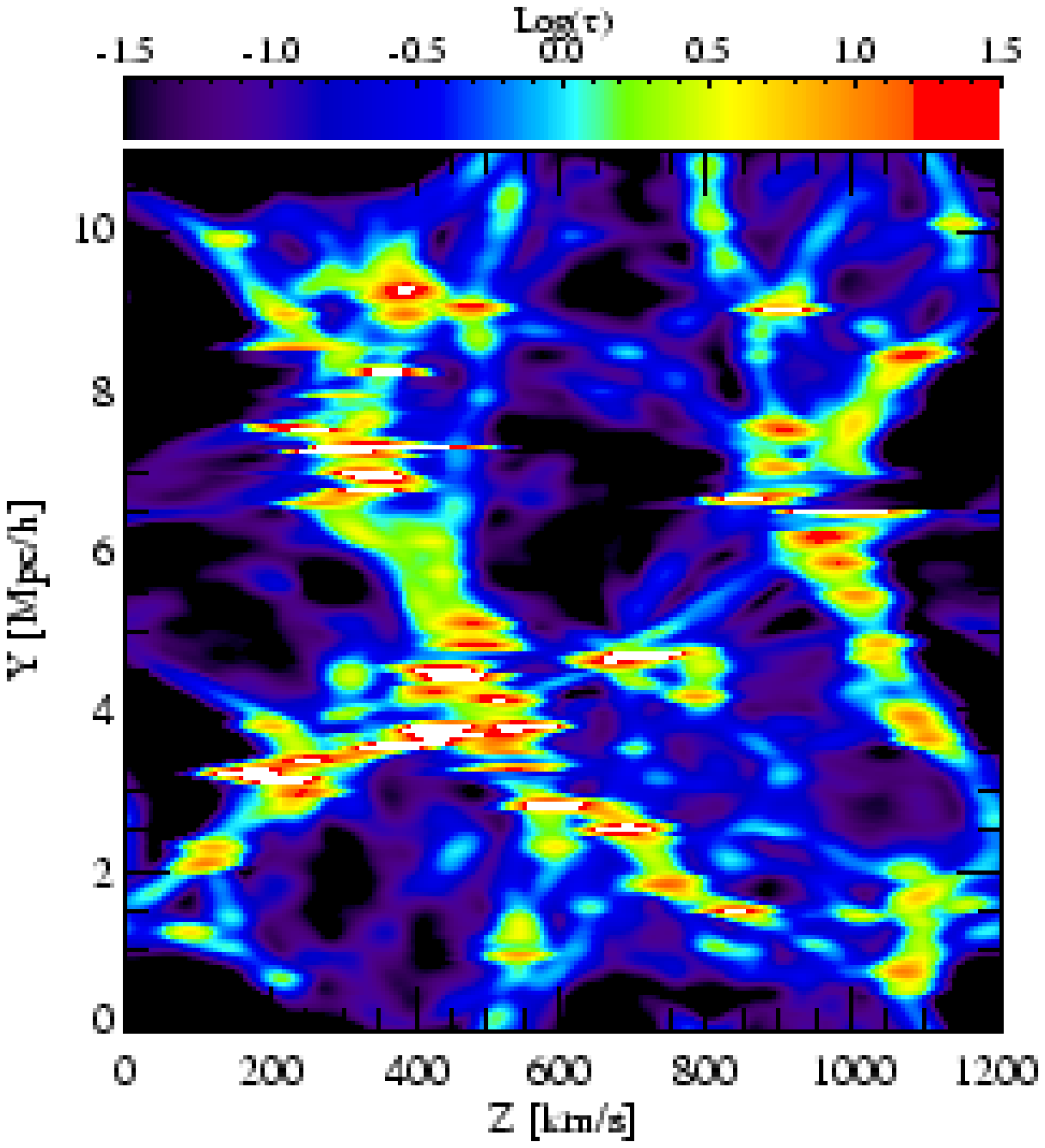}
}
\caption{2-d Optical depth distributions within the simulations.
Slices are 1 simulation cell thick, cut midway through the {\it
Left}: L11-Low, {\it Middle}: L11-Medium, and {\it Right}: L11-High
simulations.  Color codes are as indicated on top of each panel. These
optical depth maps show clearly the filament-void structure that is a
universal feature of the concordance cosmological model. The horizontal features are due to redshift-space distortions (the ``finger-of-god'' effect).}
\label{fig:wmapsheets}
\end{figure*}

\begin{figure*}
\centerline{
\epsfxsize=2.5truein
\epsfbox[14 36 377 443]{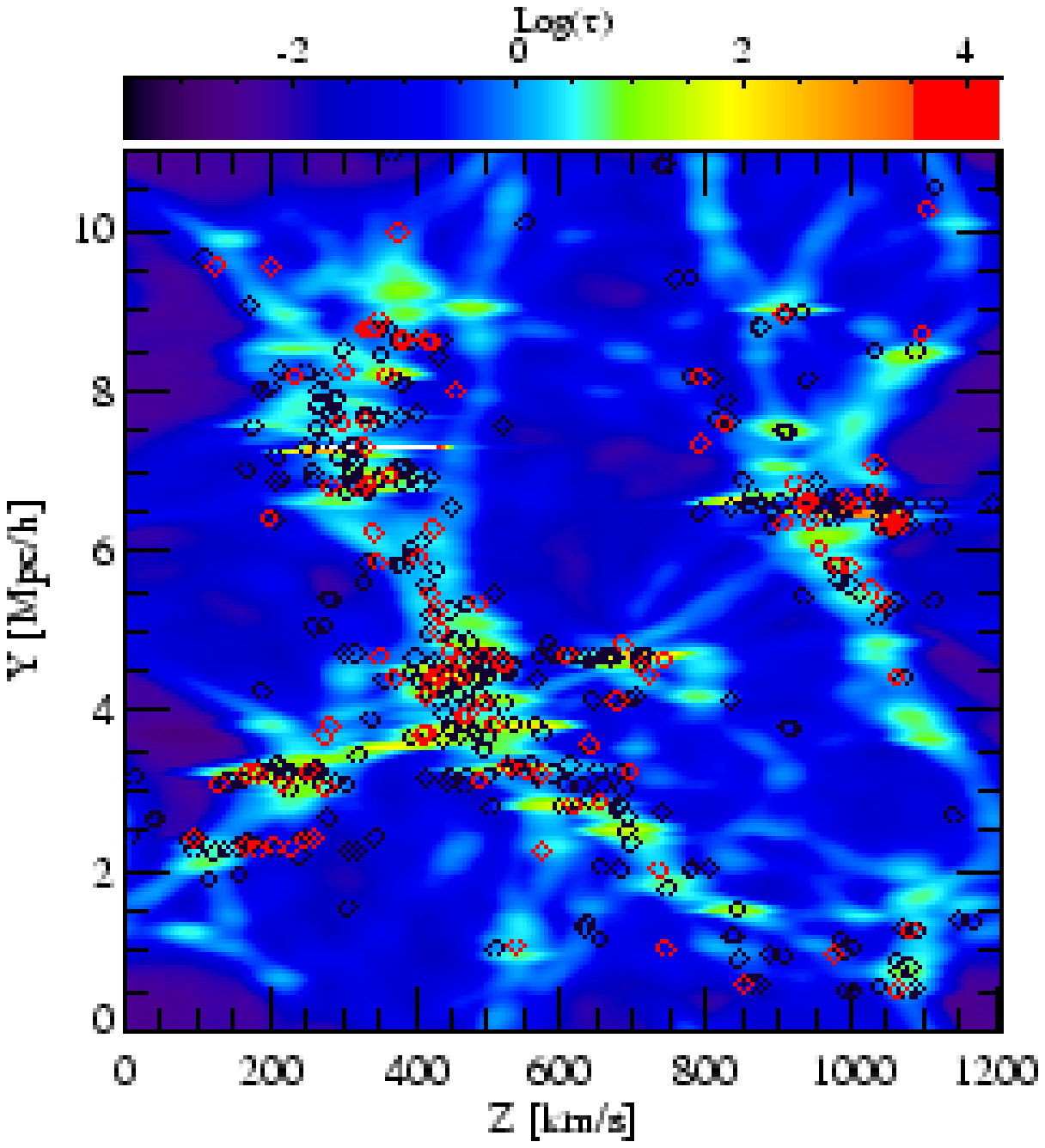}
\epsfxsize=2.5truein
\epsfbox[14 36 377 443]{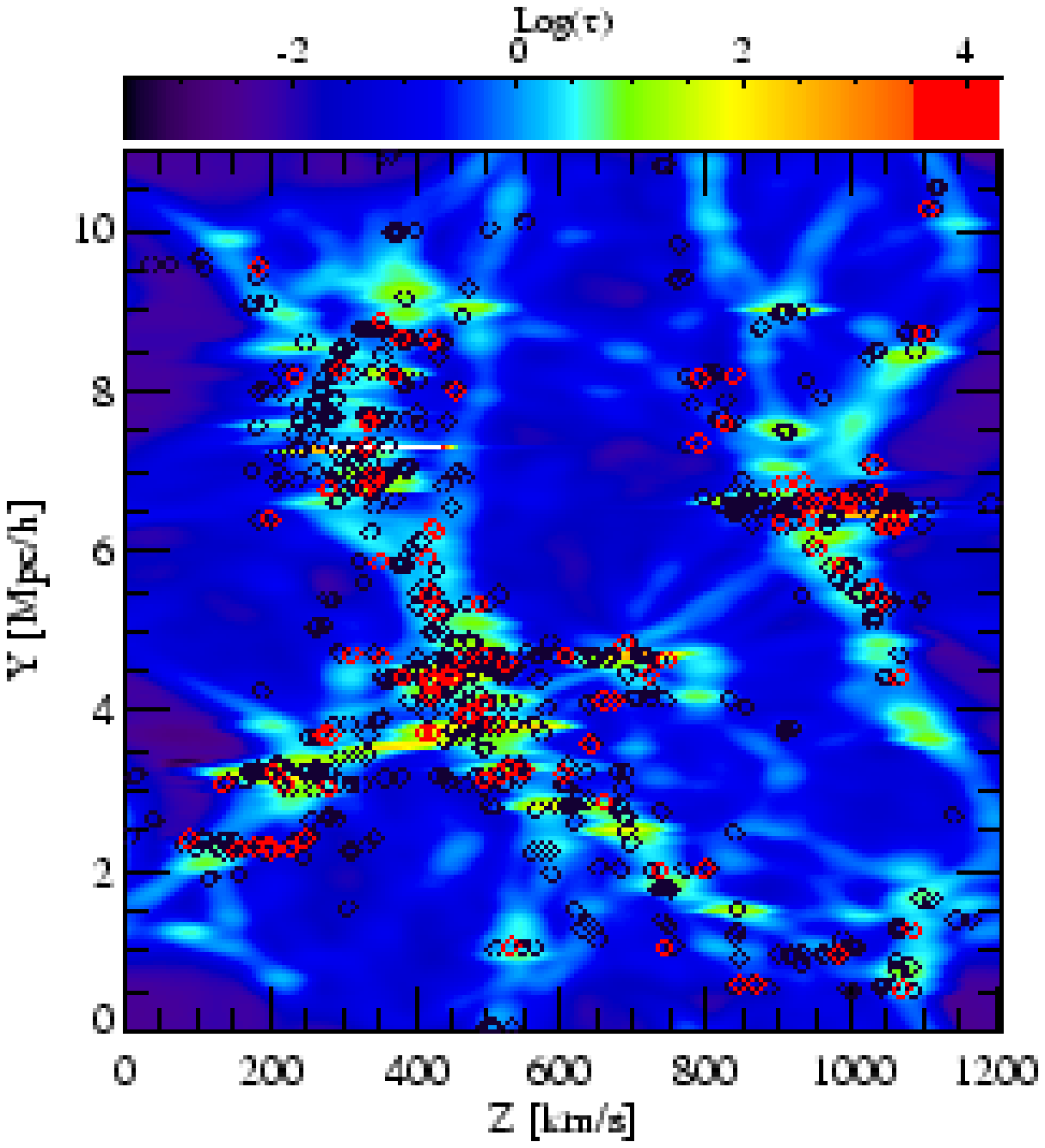}
\epsfxsize=2.5truein
\epsfbox[14 36 377 443]{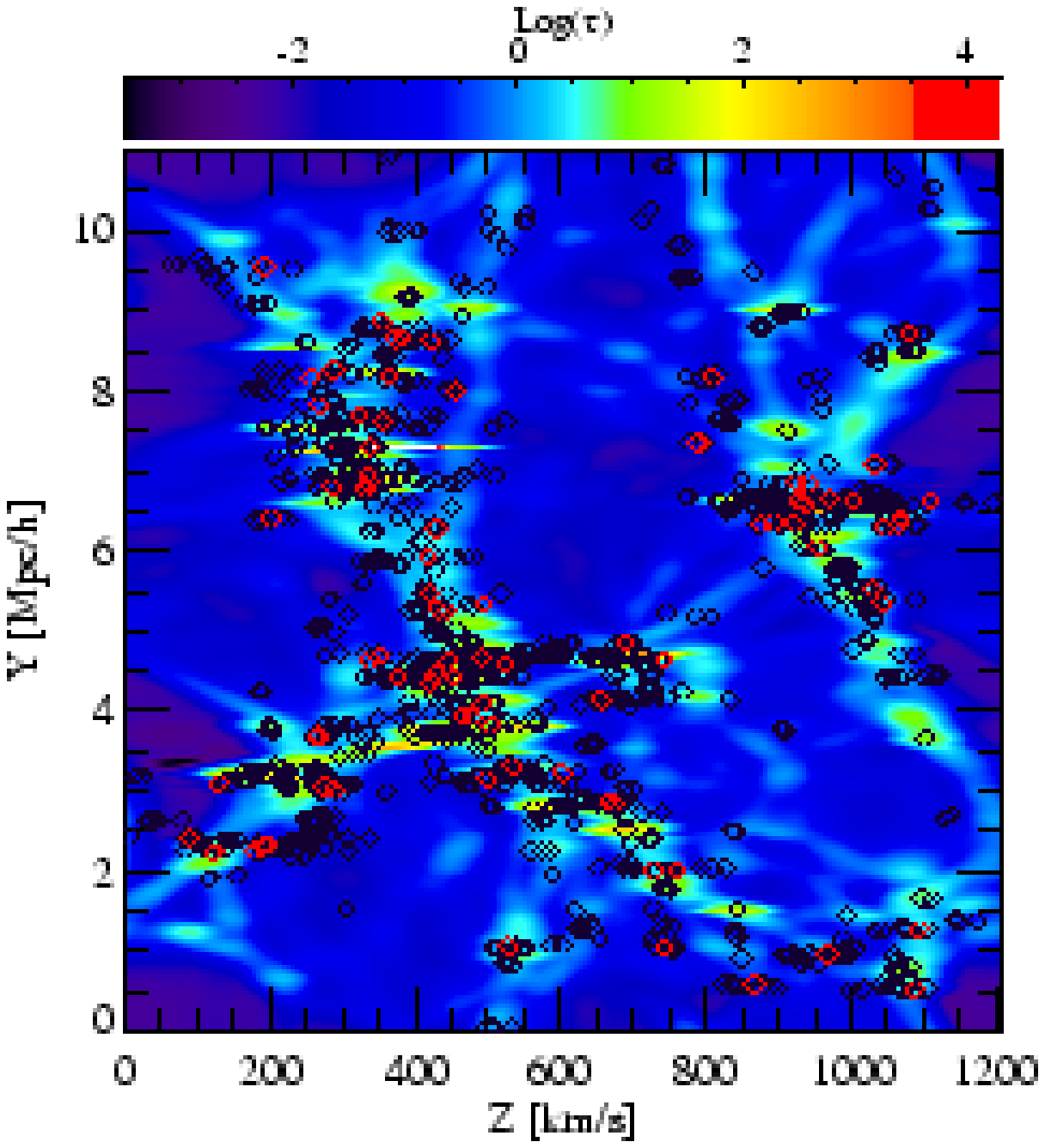}
}
\caption{Relation between galaxies and optical depth within the
simulations.  Slices are 1 simulation cell thick, cut midway through
the {\it Left}: L11-Low, {\it Middle}: L11-Medium, and {\it Right}:
L11-High simulations.  Color codes are as indicated on top of each
panel.  All galaxies within $1\hmpc$ on either side of the slice are
shown at their locations by black symbols.  In general, galaxies trace
the filaments in the dark matter, and hence the gas and optical depth,
distributions. The positions of galaxies with stellar masses larger
than $10^8\msun$ are indicated by the red symbols.}
\label{fig:wmapgalsheets}
\end{figure*}

  The global behavior of the \lya forest is more clearly seen in
Figure~\ref{fig:wmapsheets}, which shows 2-dimensional, 1-cell thick
slices through the L11-High, L11-Medium, and L11-Low simulations. We
plot the logarithm of the optical depth as a function of spatial
position along the vertical axis, and velocity along the horizontal
axis. Hence, each horizontal line in the slice shows a spectrum like
that in Figure~\ref{fig:los} (the transmitted flux shown in
Fig.~\ref{fig:los} is related to the optical depth as
$F=\exp(-\tau)$). Figure~\ref{fig:wmapgalsheets} shows the same slices
(except for a modified color coding for the optical depth), together
with the spatial positions (along the vertical axis) and redshifts
(along the horizontal axis) of galaxies in the simulation that are
located within $1\hmpc$ on either side of the sheets. Galaxies with
stellar masses larger than $10^8\msun$ are shown in red.  The
connection between galaxies and the \lya forest is clearly displayed
in Figure 6. Galaxies always lie in regions of high optical depth
(because all galaxies within $1\hmpc$ of the sheet are shown, a small
number of them appear to be displaced from high optical depth regions,
but they are all at high optical depths when they are plotted on a
slice sufficiently close to the galaxy). There is a remarkable
similarity of the optical depth slices among the three simulations,
despite the very different wind strengths that were assumed.

\begin{figure*}
\centerline{
\epsfxsize=2.5truein
\epsfbox[14 36 377 443]{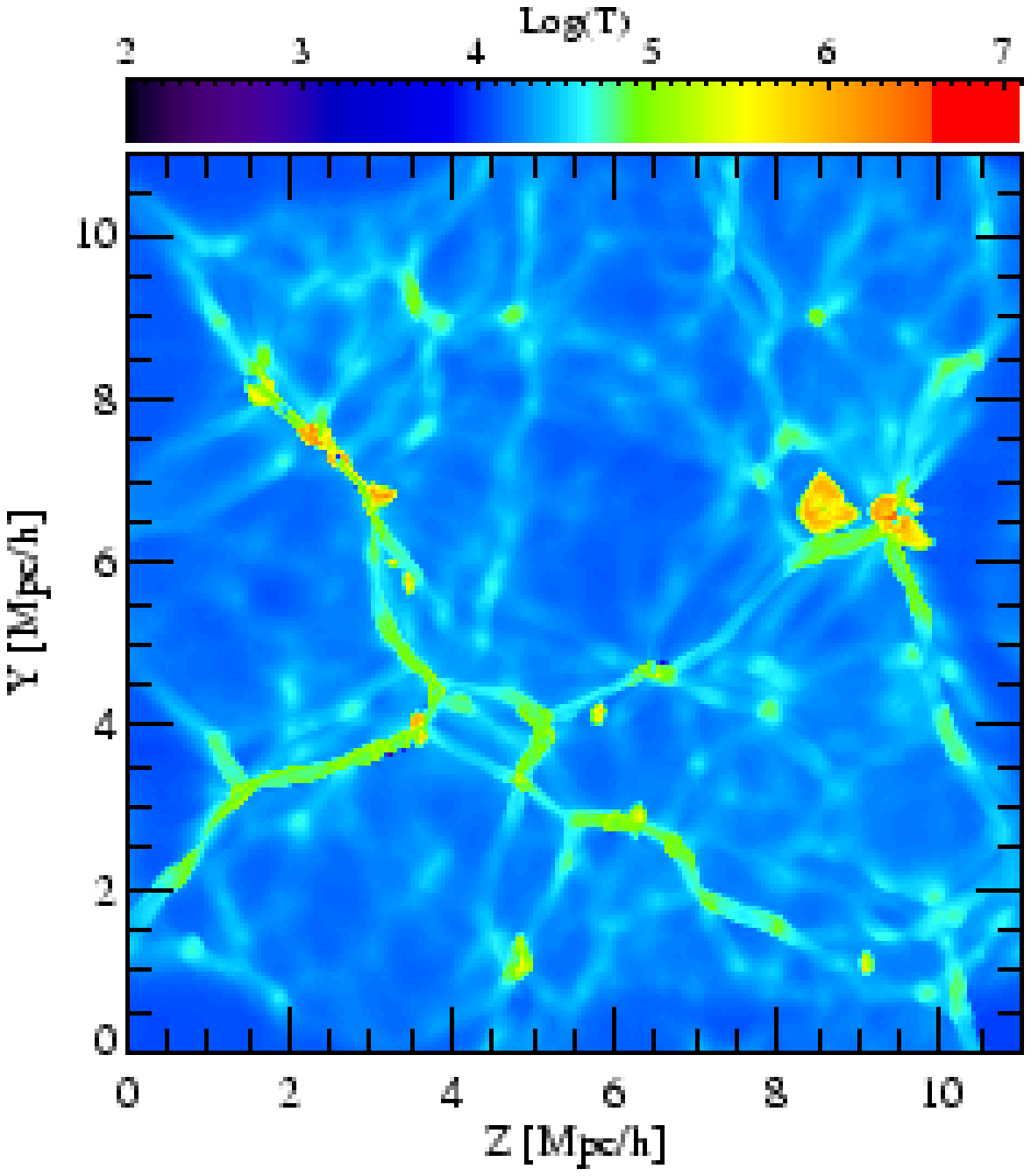}
\epsfxsize=2.5truein
\epsfbox[14 36 377 443]{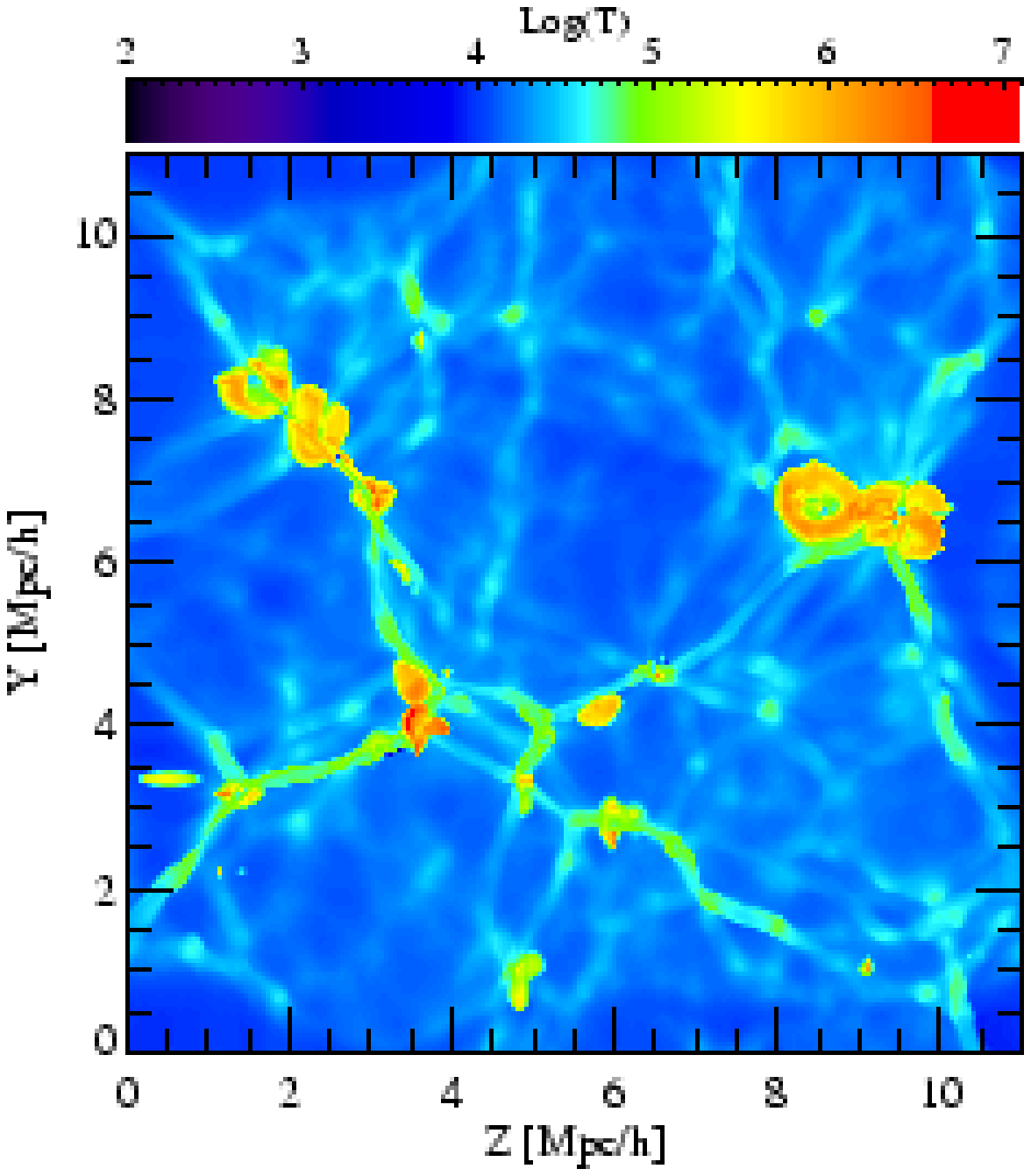}
\epsfxsize=2.5truein
\epsfbox[14 36 377 443]{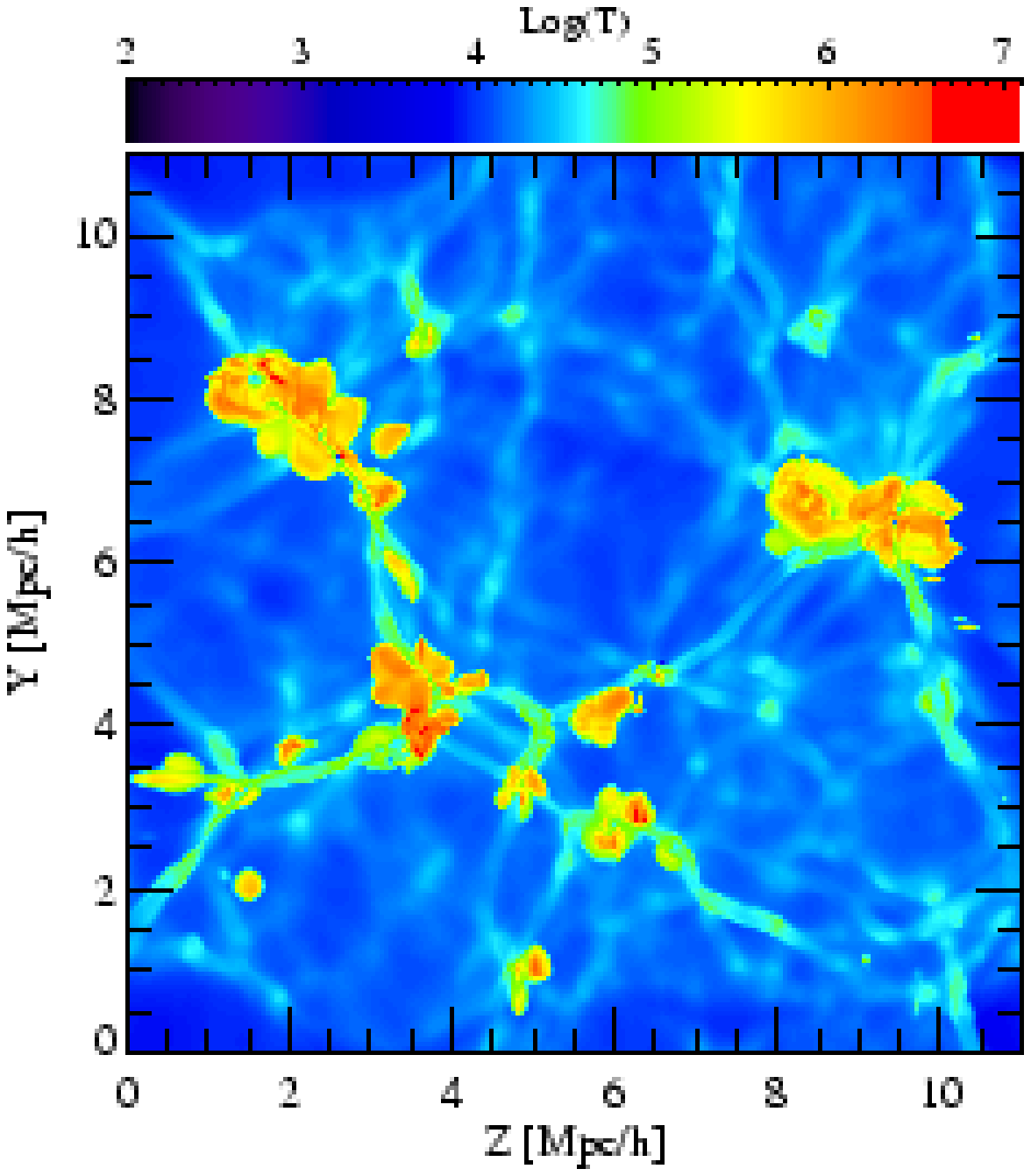}
}
\caption{2-d temperature distribution within the simulations.  Slices are now shown in log(T) through the {\it Left}: L11-Low, {\it Middle}: L11-Medium, and {\it Right}: L11-High simulations.  The presence of feedback in the temperature structure is clear in these maps.  The L11-High simulation shows large bubbles of high temperature gas surrounding ``knots'' of high density in both the galaxy and optical depth distribution.  The temperature bubbles are clearly taking the ``path of least resistance'' and expanding into the lower density medium perpendicular to the filaments.}

\label{fig:wmaptempsheets}
\end{figure*}

\begin{figure*}
\centerline{
\epsfxsize=2.5truein
\epsfbox[14 36 377 443]{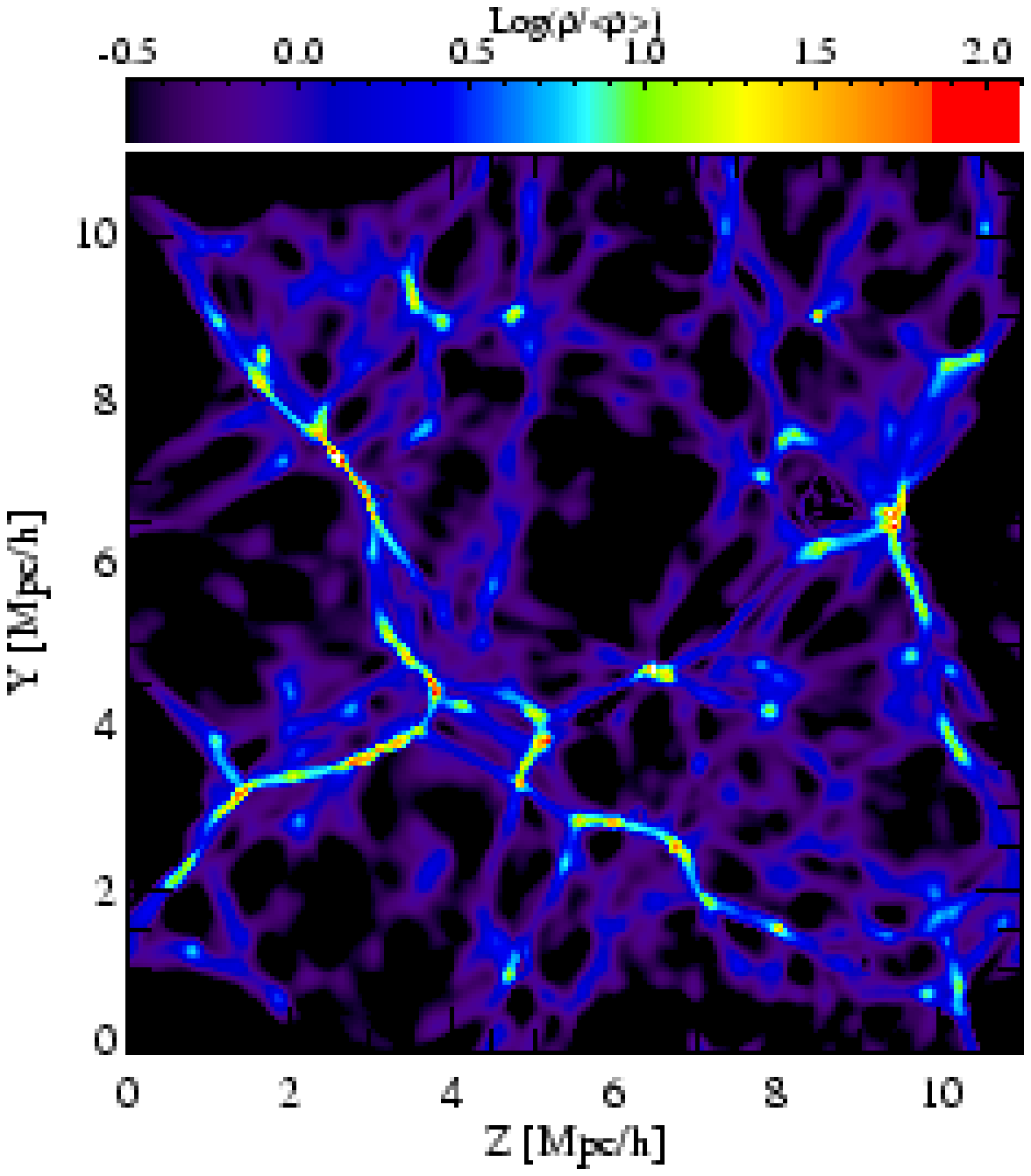}
\epsfxsize=2.5truein
\epsfbox[14 36 377 443]{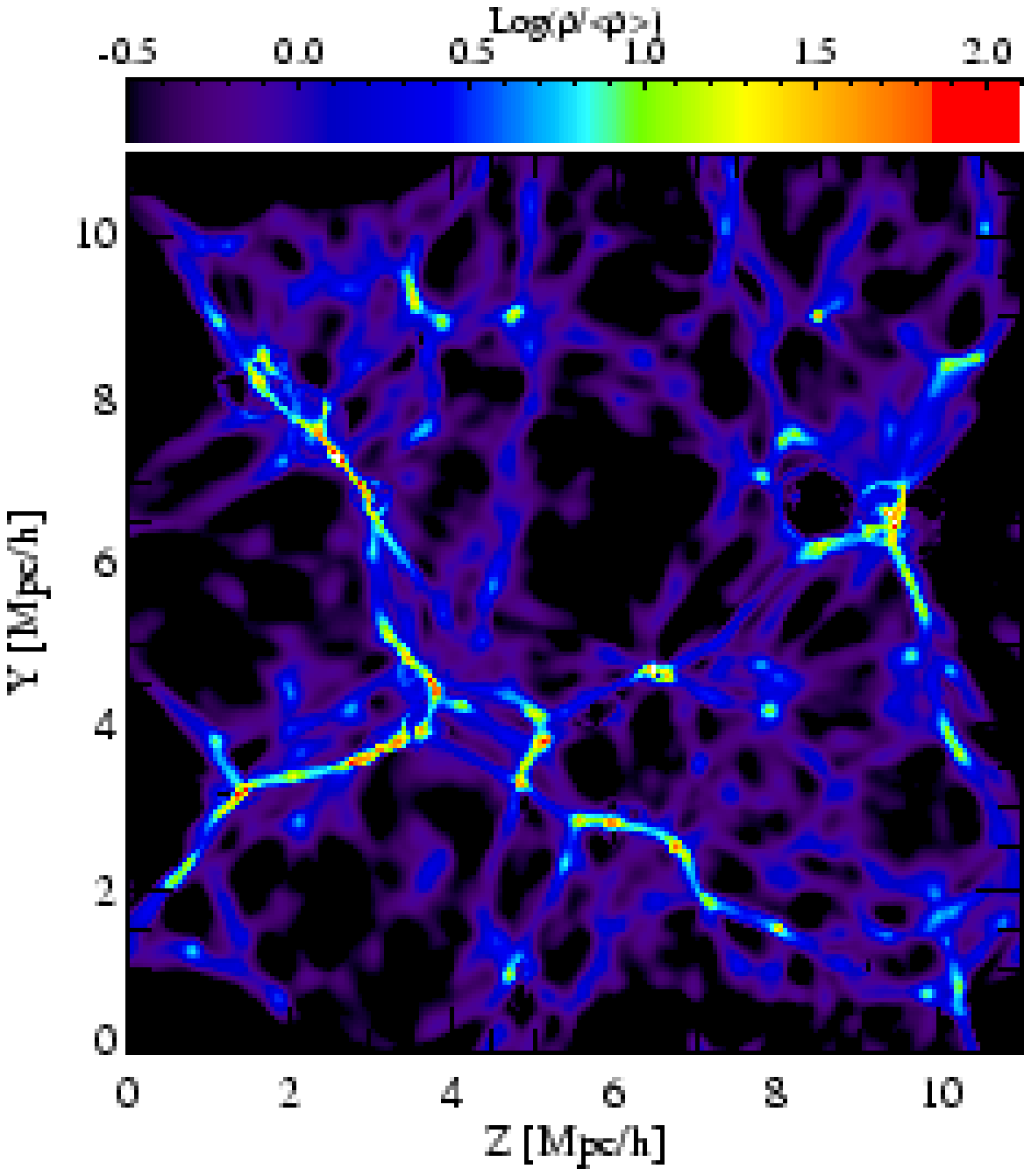}
\epsfxsize=2.5truein
\epsfbox[14 36 377 443]{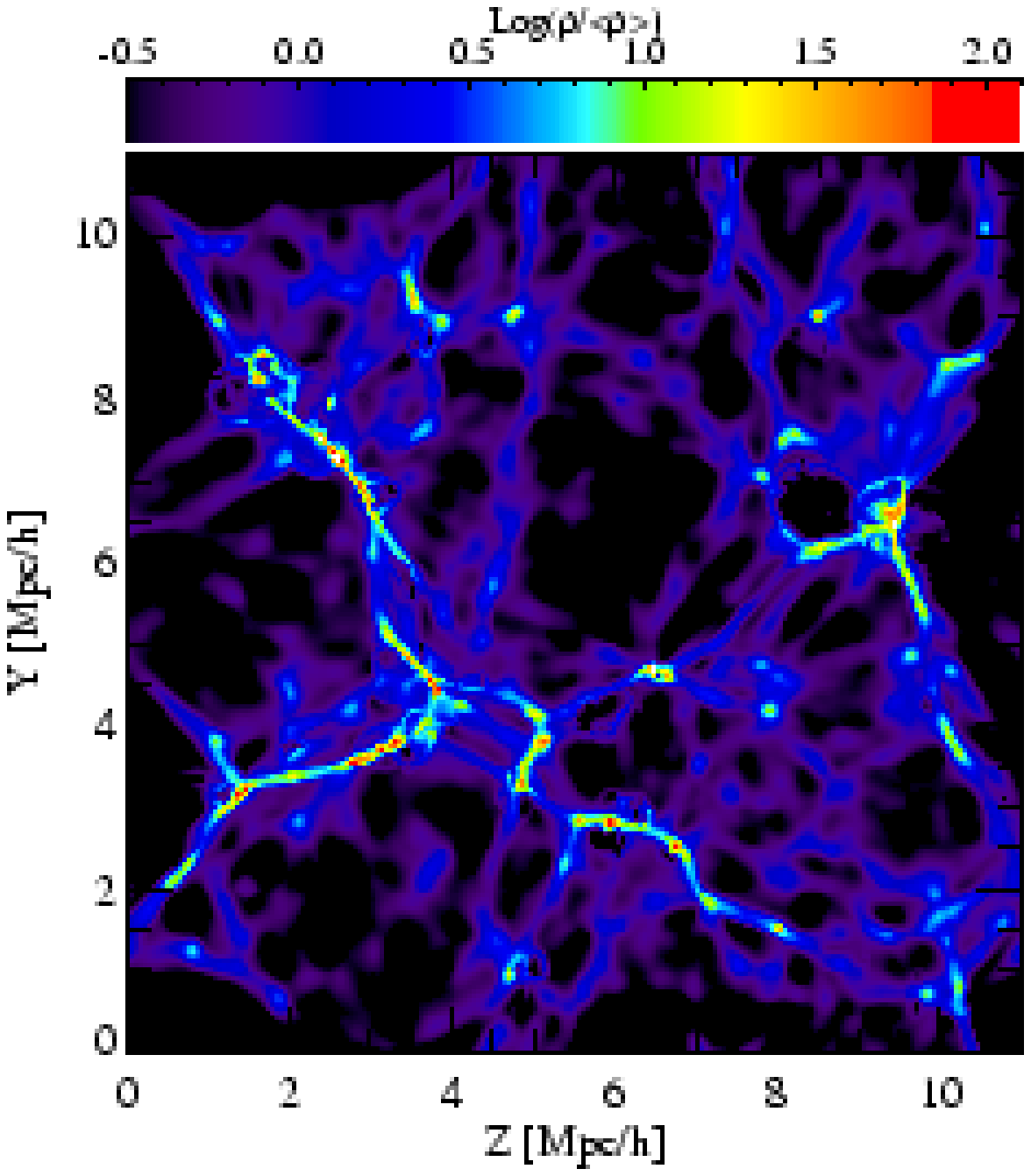}
}
\caption{2-d density distribution within the simulations.  Slices are now shown in log($\rho$/${\bar \rho}$) through the {\it Left}: L11-Low, {\it Middle}: L11-Medium, and {\it Right}: L11-High simulations. In contrast to the temperature structure, the density structure of the simulation seems remarkably {\it unaltered} by the presence of feedback. }
\label{fig:wmaprhosheets}
\end{figure*}

  To understand this result in terms of the physical properties of the
IGM, we plot the temperature and density of the same sheet in real
space for the three simulations in Figures~\ref{fig:wmaptempsheets}
and \ref{fig:wmaprhosheets}. Note that the correspondence to the
optical depth sheet shown in Figures~\ref{fig:wmapsheets} and
\ref{fig:wmapgalsheets} is not exact because the gas peculiar
velocities redistribute absorption along the line of sight. The
bubbles produced by galactic winds are very clearly seen in the
temperature structure. Similar wind structures have been found in
other simulations using the SPH technique (e.g., \citealt{theuns02,
bruscoli03}).  The size of the hot bubbles clearly increases with the
strength of the winds. However, the winds do not result in the complete
removal of high-density gas near the parent galaxies from which they
originate. In fact, Figure~\ref{fig:wmaprhosheets} shows that the gas
density structure is affected only weakly by the winds: comparing the
three simulations, we find that the filaments located near the hot
bubbles are broken up at a few places on small scales, but apart from
this they largely remain in place. The winds have some effect on the
optical depth maps of Figures~\ref{fig:wmapsheets} and
\ref{fig:wmapgalsheets}: for example, at the position $Y\simeq 8 $
Mpc/h and $Z\simeq 250 \kms$, there is a region over which \lya
absorption disappears as the wind strength is increased. This region
corresponds to the large hot bubble seen in
Figure~\ref{fig:wmaptempsheets} near this position, and the filament
in Figure~\ref{fig:wmaprhosheets} that is broken in the high-winds
simulation at the same position.

  Such strong wind effects on the optical depth are, however,
relatively rare, and when they occur they tend to redistribute the
absorption rather than eliminate it. Most of the time, the hot bubbles
expand into low density regions around galaxies, which produce little
absorption in any case (see, for example, the region at $Y\simeq 6.5$
Mpc/h, $Z\simeq 8.5$ Mpc/h where a low-density region increases in
size with wind strength in Fig.~\ref{fig:wmaprhosheets}). In other
words, the winds follow the path of least resistance.  Other gas
around galaxies may actually get compressed by the wind bubbles and
increase its \lya absorption. Voids in the \lya spectrum created by
the hot expanding bubbles of the winds tend to be filled in by
absorption from adjacent gas falling into the galaxy (owing to varying
peculiar velocities).
The net result is that the effect of winds on the \lya spectra is
rather weak and highly variable among different lines of sight.

\begin{figure*}
\centerline{
\epsfxsize=3.5truein
\epsfbox[30 239 244 711]{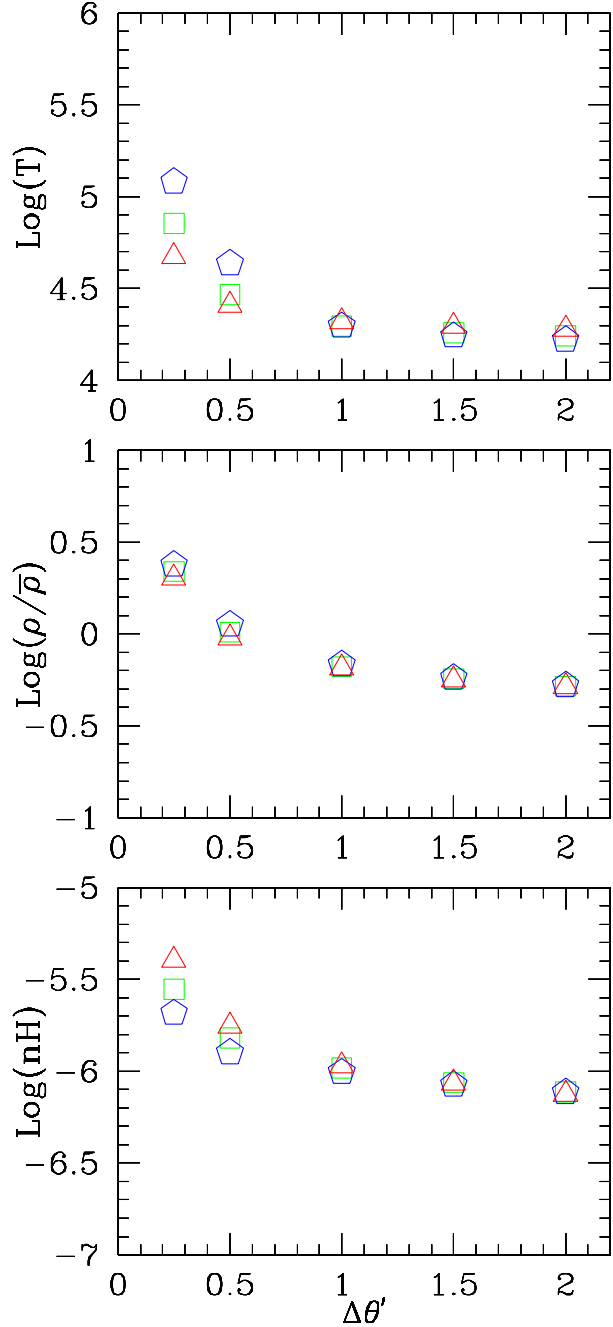}
}
\caption{Physical conditions near galaxies within the simulations as a
function of wind strength.  Median quantities for {\it Top}: Log
Temperature, {\it Middle:} Log gas overdensity, {\it Bottom:} Log
neutral density for the Low (red triangles), Med (green squares), and
High (blue pentagons) simulations. The temperature is clearly higher
in the High feedback simulation than in the Medium or Low simulations
but the density is only slightly lower.  The neutral density continues
to rise close to galaxies in all simulations, although the absolute
value is lower in the High feedback simulation due to the increased
temperature.}
\label{fig:wmapphys}
\end{figure*}

  Figure~\ref{fig:wmapphys} quantifies the impact of winds on the
physical state of the gas in the neighborhood of galaxies. We show the
median value of the temperature, gas overdensity, and neutral hydrogen
density in all the simulation pixels whose distance to any identified
galaxy is within the value on the horizontal axis. The temperature
increase and neutral density decrease with wind strength is apparent
here. Note, however, that even in the simulation with the strongest
winds the neutral density continues to rise as a galaxy is approached.
Curiously, the total gas density is not much affected by the winds; if
anything it tends to increase slightly with wind strength. These effects
of winds on the physical state of the gas are the direct cause of any
changes in the \lya forest transmission close to galaxies.

\section{Flux Statistics}

  Here we quantify the relation between the \lya forest flux
statistics and the proximity to galaxies in the simulation, and the
way this relation varies with wind strength.

\subsection{The Conditional Flux Decrement PDF}

\begin{figure*}
\centerline{
\epsfxsize=6.5truein
\epsfbox[41 286 571 699]{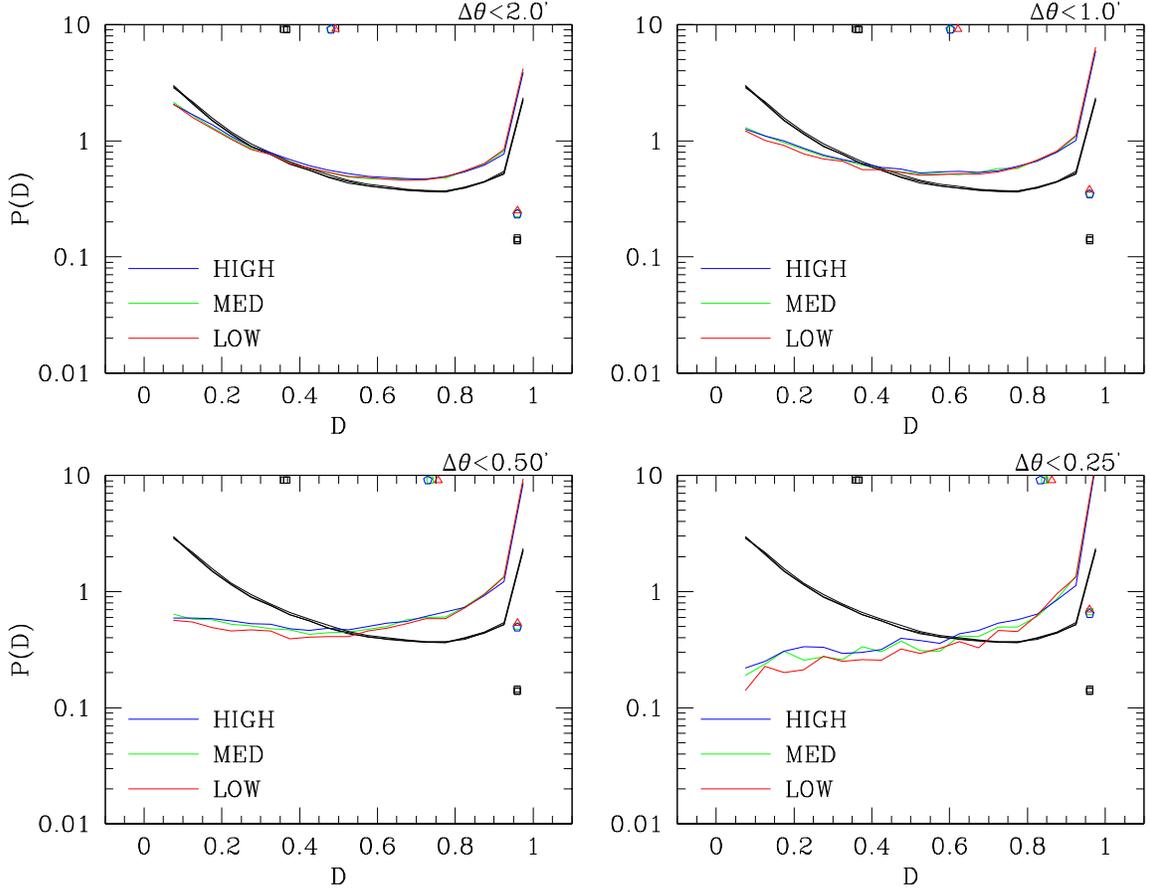}
}
\caption{Conditional flux decrement PDF as a function of wind strength.  PDFs are computed for all galaxies within the L11-Low(red), L11-Medium(green) and L11-High(blue) simulations.  Black curves show the unconditional PDF for each simulation.  Panels indicate decreasing separation between galaxies and the sightline in arcminutes as indicated in top right hand corner of each panel.  Symbols along the top of each panel show the mean of the distributions and symbols along the right hand side indicate the fraction of pixels in each distribution with flux decrement $0.9 \leq {\rm D}\leq 1.0$ for each simulation.  Red triangles, green squares and blue pentagons correspond to the L11-Low,Medium and High simulations respectively. The conditional flux PDFs are remarkably insensitive to galactic winds.}
\label{fig:wmappdfall}
\end{figure*}

  In general, one can codify the influence of a galaxy on the \lya
forest by the probability distribution function (PDF) of the flux
decrement, $D$ (equal to one minus the fraction of transmitted flux),
conditioned on the presence of a galaxy within a transverse angular 
separation $\Delta\theta$ and velocity separation $\Delta v$. We first study this
conditional flux PDF (CPDF) for $\Delta v=0$, which was studied
previously by \cite{kollmeier03a} using SPH simulations and is
described in more detail there. We obtain the CPDF using all resolved
galaxies (defined as having at least 20 star particle members) in each
simulation. For each galaxy, random lines of sight that pass within a
maximum impact parameter $\Delta\theta$ are chosen, and the flux
decrement at the pixel that has the same redshift as the galaxy
(taking into account the galaxy peculiar velocity) is selected. Note
that by choosing lines of sight uniformly distributed within a maximum
impact parameter, we are averaging the CPDF over impact parameters
with a weighting proportional to $\Delta\theta\, d\Delta\theta$, up to the maximum
impact parameter shown in each panel. The CPDF of the flux decrement
is shown in Figure~\ref{fig:wmappdfall} for the four values of
$\Delta\theta$ (in arcminutes) indicated in the top right corner of
each panel. The results are shown for the three L11 simulations with
different wind strengths. The black solid lines show the unconditional
PDF computed using all pixels in all spectra.

  As the transverse galaxy-sightline distance is reduced, there is a
shift in the CPDF from unsaturated (low D) pixels to saturated
(high D) pixels. The values of the mean decrement (shown as the top
symbols) and the probability of saturated pixels, defined as $D>0.9$
(right symbols), which capture some of the information of the full
CPDF, also show this clearly. This result simply reflects the
association of galaxies with regions of high optical depth shown
earlier in Figure 6 and is in good agreement with the results of
\cite{kollmeier03b}.

  The wind effects are seen to be small, as expected from the optical
depth sheets discussed in the previous section. These weak wind
effects on the \lya forest are in agreement with the work of
\cite{theuns02} and \cite{bruscoli03} using a different simulation
method with different wind prescriptions.  The result also agrees with
that of \cite{croft02,kollmeier03b,weinberg03} and
\cite{desjacques04}, who examined simple ``bubble'' wind models. The
small difference has the expected sign: stronger winds should and do
reduce the \lya absorption in the regions from which they emerge.

\begin{figure*}
\centerline{
\epsfxsize=6.5truein
\epsfbox[41 286 571 699]{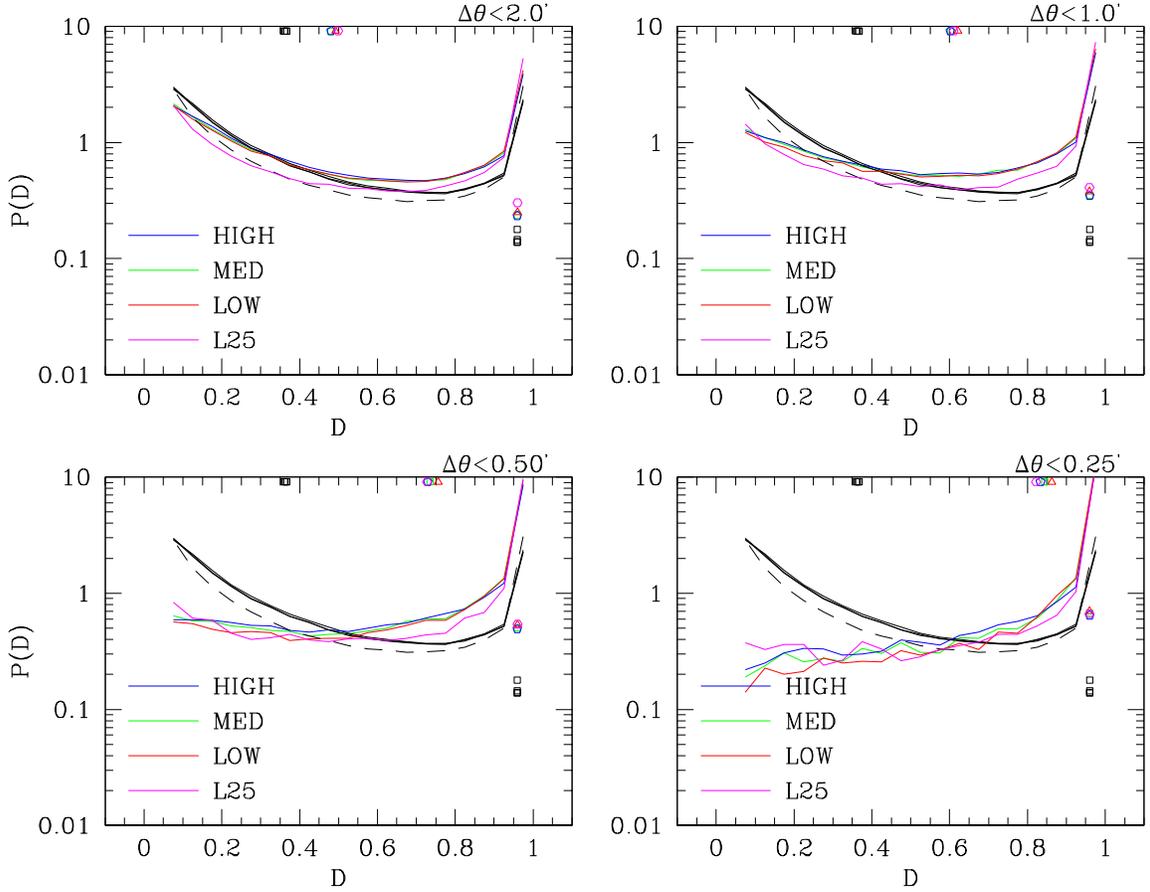}
}
\caption{Effect of box size on conditional flux decrement PDF.  PDFs are
computed for all galaxies within the L11-Low(red), L11-Medium(green),
L11-High(blue) and L25 (magenta) simulations.  Panels indicate
decreasing separation between galaxies and the sightline in arcminutes
as indicated in top right hand corner of each panel.  The
unconditional flux decrement is shown for the L11 simulations (black
solid lines) and the L25 simulation (black dashed lines).  Symbols
along the top of each panel show the mean of the distributions and
symbols along the right hand side indicated the fraction of pixels in
each distribution with flux decrement $0.9 \leq {\rm D}\leq 1.0$ for
each simulation.  Red triangles, green squares and blue pentagons
correspond to the L11-Low,Medium and High simulations
respectively. Magenta hexagons correspond to the L25 simulation.  The
L25 simulation has different unconditional flux PDF due to the larger
scale modes that are allowed in this box.  The conditional flux PDFs,
however, are at the expected level of agreement given the differences
in box size and parameters between the four simulations.}
\label{fig:comppdf}
\end{figure*}

  In Figure~\ref{fig:comppdf}, we compare the PDFs for all four of our
simulations. The small difference in the unconditional flux PDFs
between the L25 simulation (dashed black line) and the three L11
simulations (solid black lines) is due to a combination of the effect
of differing box size and the slightly different power spectra adopted
in the large and small boxes. Simulations of small boxes have an
artificially low dispersion in the flux decrement that arises from a
low dispersion in gas density distribution, owing to the suppression
of large-scale fluctuations.

  In summary, the unconditional and conditional PDF of the flux
decrement are robust predictions of the CDM model, and the effects of
galactic winds on them are small.

\subsection{Dependence of the Conditional PDF on Galaxy Luminosity}

  The conditional PDF can also be examined as a function of galaxy
luminosity to see whether more luminous galaxies tend to be associated
with stronger \lya absorption. In \cite{kollmeier03b}, variations of
\lya absorption around galaxies were shown to have little dependence
on the total star formation rate in each galaxy. The Eulerian
simulations we use here have a different prescription for estimating
galaxy luminosity (see \S 2), and they incorporate the effects of
winds, which can affect galaxies of different luminosities in different
ways.

\begin{figure*}
\centerline{
\epsfxsize=6.5truein
\epsfxsize=6.5truein
\epsfbox[41 286 571 699]{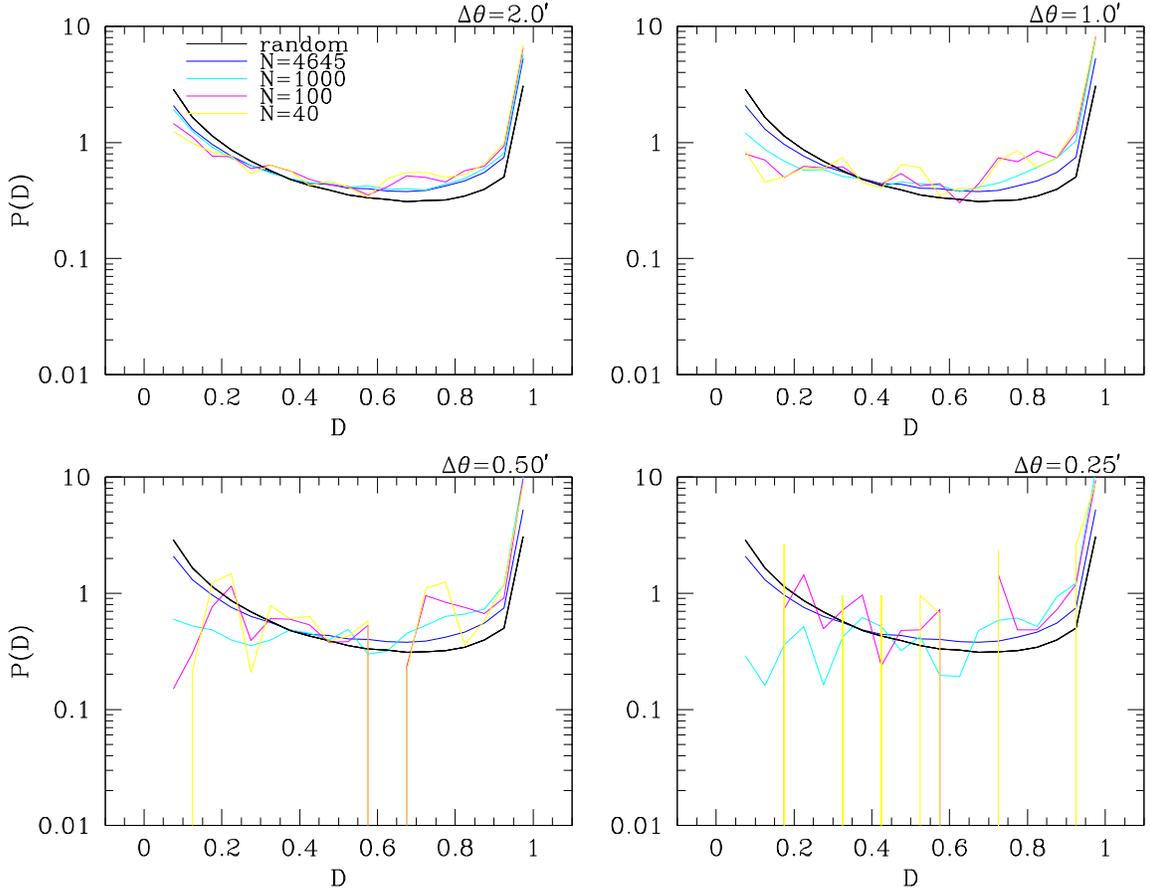}
}
\caption{Effect of galaxy luminosity on conditional flux decrement
PDFs.  Curves show different galaxy samples selected by increasing
thresholds in B-band luminosities.  The calculation computed around
all galaxies is shown in blue, the 1000 brightest galaxies (cyan), the
100 brightest galaxies (magenta), and the 40 brightest galaxies
(yellow).  Panels show decreasing maximum angular separation
$\Delta\theta^{\prime}$ (indicated in the upper right hand corner of
each panel).  The yellow curve in each panel shows the brightest
galaxies whose number density matches that of observed LBGs
\citep{adelberger98,adelberger04}.  The expected trend of increased
saturated fraction with decreasing sightline-galaxy separation is
clear. However, there is no strong dependence of the PDF on
luminosity.  The paucity of galaxy-sightline pairs make the
closest-separation, smallest-number curves noisy.}
\label{fig:l25pdf}
\end{figure*}

  We explore the dependence on galaxy luminosity in
Figure~\ref{fig:l25pdf}, which shows the CPDF in the L25 simulation
within four different impact parameters and for galaxies with
luminosity above four different thresholds. The four luminosity
thresholds (with luminosities computed as described in \S 2.2) are
$7.124\times 10^9$, $3.358\times 10^9$, $2.809\times 10^8$, and
$9.102\times 10^5\lsun$, and the total number of galaxies in the $25
\hmpc$ box of the simulation above each luminosity threshold is
labeled for each curve. For reference, the space density of the
observed LBGs would correspond to our highest luminosity threshold,
with only $N=40$ galaxies in the simulation. The figure shows there is
a weak dependence of the galaxy-\lya transmission correlation on
galaxy luminosity.  The sign of the effect is as expected if galaxy
luminosity correlates with mass and large-scale environment---the \lya
forest around more luminous galaxies shows a preference for high
decrement pixels.  This dependence on galaxy luminosity is, of course,
sensitive to the relation between observed galaxies and galaxies in
the simulation, which is affected by our poor understanding of the
factors that determine star formation rates in galaxies.  At the
smallest angular separations, the dependence of the galaxy-\lya
transmission correlation cannot be clearly distinguished from the
random noise on the curves caused by the small number of
galaxy-sightline pairs at these separations in our box.  It is clear,
however, that in all cases the \lya absorption increases as a galaxy
is approached.

\subsection{Dependence on Velocity Offset}

  The preceding figures show only the CPDF of the flux decrement at
the same velocity as a galaxy within a certain impact
parameter. However, we expect the galaxy-intergalactic medium
connection to persist over some velocity range from the true redshift
of the galaxy.  One can choose to examine pixels within a fixed
velocity interval from the galaxy redshift as a further test of the
predictions.  We now examine the flux distribution as a function of
velocity offset.

  In Figure~\ref{fig:wmappflux} we show the median, first quartile and 10 percentile
values of the transmitted flux, F, for \lya forest spectral pixels in lines of
sight within annuli of angular separation from a galaxy in the range
$\theta_1 < \Delta\theta < \theta_2$, and within a velocity interval
$\pm\Delta v$. The 100 most massive galaxies in the simulation have
been used for this calculation.  Hence, the (top, middle, bottom)
curve in each panel gives the value of the transmitted flux for which
the probability to measure a larger flux is $0.1$, $0.25$, and $0.5$,
respectively, over the indicated range of $\Delta\theta$ and at any pixel in the spectrum within $\pm\Delta v$. The six panels are for six different ranges of
$\Delta\theta$, as shown. The red, green and blue curves correspond to
the L11-Low, L11-Medium and L11-High simulations, respectively. In the
limit of large $\Delta v$, the curves generally approach the
unconditional percentile values. The decrease of $F$ at small $\Delta
v$ shows that it becomes increasingly unlikely to observe large
transmitted fluxes as one probes pixels closer to the galaxy redshift.

  We note from Figure~\ref{fig:wmappflux} that the effects of winds
are most important within an impact parameter $\Delta\theta < 0.25'$
and that they are small at larger separations. We also point out that
the probability of observing a transmitted flux value as high as $\sim
0.8$ near a galaxy decreases rapidly only at $\Delta v \lesssim 150
\kms$, and that at larger velocity separations this probability is not
much affected by the presence of winds (see top left panel).  In other
words, close to the galaxy's systemic velocity, the probability of
finding a transparent line of sight is low but is significantly
increased by winds, while at large offsets from the galaxy redshift
the probability is higher but is no longer sensitive to winds.  The
median transmitted flux, typically with a lower value $\sim 0.1$ to
$0.6$, is more strongly affected by winds up to velocity separations
$\Delta v \sim 300 \kms$. As seen in Figure~\ref{fig:wmappflux}, the
effects of winds should be detectable in the \lya forest near galaxies
once the median transmitted flux is measured to an accuracy better
than $\sim 10\%$.

  The results obtained from the L11 simulations are, however, strongly
affected by the size of the simulation box, and better theoretical predictions from larger boxes will be needed in order to infer wind strengths from the transmitted flux distribution as a function of $\Delta v$ and $\Delta \theta$. The total velocity across
the box in the L11 simulations is only $\sim 1200\kms$, so the
velocity interval $\Delta v=300\kms$ shown in
Figure~\ref{fig:wmappflux} is nearly half of the box size.  The curves in
this figure are strongly affected by the box size: they increase too
fast with $\Delta v$ due to the suppression of large-scale power.  Any
box-size effects should be smaller in the L25 simulation. The same
curves for this simulation, computed for the set of the 100 most
luminous galaxies, are shown in Figure~\ref{fig:pfluxl25}. These
curves are more reliable predictions for a direct comparison to
observations than those in Figure~\ref{fig:wmappflux}. Unfortunately,
we do not have simulations for different wind strengths in larger
boxes.  However we expect that the differences introduced by winds
would be similar to those in the L11 simulations.

\section{Comparison with Observations}

  The correlation of LBGs and the \lya forest has been studied
observationally by \cite{adelberger03}, who measured the mean flux
decrement ($\langle D\rangle =\langle 1- F\rangle$, where $F$ is the
transmitted flux, which was shown in Figs.~\ref{fig:wmappflux} and
\ref{fig:pfluxl25}) as a function of total redshift-space comoving
separation from a galaxy, $\Delta$. This is defined in terms of the
angular and velocity separations as
\begin{equation}
\Delta^2 = \left[ D_a(z) (1+z) \Delta\theta \right]^2 +
\left[ (1+z) \Delta v / H(z) \right]^2 ~,
\label{eq:delta}
\end{equation}
where $D_a(z)$ and $H(z)$ are the angular diameter distance and the
Hubble parameter at redshift $z$. The \cite{adelberger03} observations
are reproduced in Figure~\ref{fig:adelcomp}, together with our
predictions for the same quantity using the L25 simulation. We use four
different galaxy luminosity thresholds, such that the total number of
galaxies in the simulation above the threshold is 1000, 100, 40 and 20.

  As previously found by several authors, the results of
\cite{adelberger03} do not agree with theoretical expectations.  The
observations show the flux decrement at a constant value for $\Delta <
3 \hmpc$, which then drops sharply at $\Delta < 0.5 \hmpc$.  The
predictions show that the flux decrement should continually increase
as a galaxy is approached. The effects of winds in our simulations do
not alter the clear increase of the flux decrement, relative to its
mean value, as $\Delta$ decreases from $3 \hmpc$ to $0.5 \hmpc$, and
they may produce at most a $\sim 10\%$ decline of the flux decrement
when $\Delta < 0.5 \hmpc$ for the most luminous galaxies. However,
this small decline, seen in the innermost point in
Figure~\ref{fig:adelcomp} for the most luminous galaxy class, may be
affected by noise due to the small number of galaxy-quasar pairs at
this small separation in the simulation.  The observational points at
small $\Delta$ in Figure~\ref{fig:adelcomp} challenge the CDM-based
picture of the Lyman-alpha forest that has emerged over the last
decade.  This picture predicts that the average flux decrement near
galaxies should increase, not decrease.  The galactic winds predicted
by our simulations reduce the strength of this trend but do not
reverse it, in agreement with other studies
\citep{croft02,kollmeier03a, kollmeier03b, bruscoli03, desjacques04}.

  We now ask whether the discrepancy seen in Figure~\ref{fig:adelcomp}
could be explained by galaxy redshift errors, misestimation of the
continuum, or simply by small number statistics.  We focus on the
innermost point, based on three quasar-galaxy pairs with separations
of $16''$, $17''$, and $21''$, for which \cite{adelberger03} find a
mean flux decrement of 0.11.  We calculate the probability of
obtaining this measured value, given the model of the L25 simulation
we have analyzed in previous sections.

%It is worth noting that the
%observational error bars in Figure~\ref{fig:adelcomp} were obtained by
%a data shuffling technique, and that to test a well defined
%theoretical model it is most appropriate to ignore the error bar and
%simply find the probability of obtaining the measured value given the
%model.

  Redshifts of faint LBGs are notoriously difficult to obtain because
the gas being expelled in a wind at small radius can displace the
redshift of emission or absorption lines relative to the systemic
galaxy redshift. After correcting for this effect, \cite{adelberger03}
estimated their galaxy redshift errors were $\sigma_v \simeq$ $150
\kms$ (according to a calibration of the redshift error that
\cite{adelberger03} measured from an independent sample of 27 Lyman
break galaxies with well measured nebular emission line redshifts),
but could be larger for some galaxies when line equivalent widths are
not measured, or in some pathological cases where galaxy wind effects
are apparently stronger. Previous analyses \citep{croft02,
kollmeier03a} found that errors in the galaxy redshifts drawn from a
Gaussian with dispersion $300 \kms$ substantially reduce the predicted
transmitted flux close to galaxies, but are insufficient to explain
the very low mean flux decrement ($\langle D\rangle \sim 0.1$) that
\cite{adelberger03} measured at the smallest value of $\Delta$ (see
Fig.\ \ref{fig:adelcomp}).  To estimate this effect within the context
of the present simulations, we calculate the cumulative distribution
of predicted decrements, $D$, averaged over velocity widths $\pm
\Delta v$, assuming various different velocity errors, $\sigma_v$.
That is, in our simulation, we make our average decrement measurement
over a velocity range $\pm \Delta v$, centered at a redshift displaced
from the true redshift of the galaxy by a random amount drawn from a
Gaussian of dispersion $\sigma_v$.  Here we use the L25 simulation
because of its larger volume.  Figure~\ref{fig:fluke} shows our
results for several choices of the averaging interval $\Delta v$.
\citet{adelberger03} binned their data by $\Delta v\sim (41, 39,
24)\,\rm km\,s^{-1}$, for the quasars SSA22D13, Q1422+2309b and
Q0201+1120 respectively. These values are bracketed by the top and
second rows of panels.  Right hand panels show the cumulative
probability distributions for a single galaxy-quasar pair, while left
hand panels show the distributions for the mean decrement averaged
over three galaxy-quasar pairs selected at random from our full
sample.

Note that while we average over velocity bins of fixed width,
\citet{adelberger03} average over a variable width that depends on
impact parameter.  However, since the range of their velocity width is
small and bracketed by the upper and middle panels in our
Figure~\ref{fig:fluke} that are essentially identical, this difference
has no practical impact.

While redshift errors substantially increase the probability of low
decrements, even for $\sigma_v=300\kms$ the probability of a single
galaxy-quasar pair having decrement $D<0.11$ is only $\sim 15\%$ for
an averaging interval $\Delta v = 50\kms$ (middle right panel), and
the probability of the mean decrement of three such pairs being below
0.11 is $\sim 1\%$.  We note that for at least one of the three
galaxies, the redshift error has been confirmed to be less than
$60\kms$ (K. Adelberger, private communication).  These probabilities
are higher than obtained in windless simulations \citep{kollmeier03a},
indicating that winds affect the flux distribution, but not enough to
make the observations likely.

In addition to an observational redshift error, there may also be an
intrinsic redshift error arising from the fact that real galaxies
may have an intrinsic velocity dispersion relative to the surrounding
gas at the observed impact parameters that is not fully included in
our simulations.  The resulting intrinsic redshift error needs to be
added in quadrature to the observational one.  In galaxy groups, the
group velocity dispersion has the same effect as a redshift error. The
simulations we use have limited spatial resolution, implying that the
high-density central parts of halos are not well resolved, and the
orbits of satellite galaxies and their tidal disruption are not
adequately followed. This results in an artificial increase of the
merger rate of galaxies in groups, which can reduce the velocity
dispersion of simulated galaxies compared to real galaxies. The
magnitude of this theoretical redshift error actually depends on the
nature of the Lyman break galaxies. If these are often satellite
galaxies undergoing intense starbursts after a group merger (e.g.,
\citealt{somerville01}), they may be moving close to the escape
velocity of the galaxy group halo and have large peculiar velocities
relative to the surrounding gas. If they are more often associated
with central halo galaxies, the peculiar velocities may be lower. For
a singular isothermal halo with velocity dispersion $\sigma$ and outer
radius $r_h$, the escape velocity at radius $r$ is
\begin{equation}
v_{esc} = 2\sigma \sqrt{1 + \log(r/r_h) } ~.
\end{equation}
For a typical galaxy group halo at $z\sim 3$ with $\sigma \sim 200
\kms$, the escape velocity may easily be as large as $600 \kms$, which
after projection may give a velocity dispersion comparable in
magnitude to the observational errors.  In other words, the total
redshift error may be substantially increased relative to the
observational redshift error.  Even f we assume, however, total
redshift errors of $600 \kms$ (which we view as unrealistically
large), the joint probability for three quasar-galaxy sightlines is
only $\sim 4\%$.  Alternatively, the high stellar masses estimated for
LBGs \citep{shapley01} suggest that they are most likely the central
galaxies of their parent halos, in which case their peculiar
velocities relative to the halo center-of-mass are likely to be small
and the intrinsic redshift errors, also small.  Observations of the
clustering of LBGs in redshift space and the pairwise velocity
dispersion of LBGs as a function of angular separation should reflect
the intrinsic dispersion of galaxy redshifts from the redshift of the
surrounding gaseous halos distinguishing between the two possible
scenarios (central vs. satellite galaxies) and constraining our
estimates of the size of this effect.

Continuum fitting errors could affect the observational estimates of
the mean decrement.  At the redshift of the \cite{adelberger03}
observations, the minimum flux decrement over intervals of $\sim 3000
\kms$, comparable to the intervals used for continuum fitting, is
usually no more than $\sim 2\%$, so it is unlikely that the continuum
fitting error is larger than that.  Spectral noise can also affect the
probability of measuring the low flux decrement of
\cite{adelberger03}. The noise in their spectrum over an interval
$\Delta v= 20 \kms$ is only $\sim 0.02$. The true flux decrement in
the first data point of Figure~\ref{fig:adelcomp} may be as high as
$0.15$, owing to the combination of continuum fitting errors and
spectral noise. The probability to measure this value of the flux
decrement for three galaxies, and for a total redshift error of
$\sigma_v=300\kms$ would still be $\sim 2\%$, from our
Figure~\ref{fig:fluke}.
 
 An alternative test that can be done to compare the model predictions
with present and future data is to average the flux decrement over a
larger interval $\Delta v$. This has the advantage of reducing the
sample variance of the measured flux decrement, and the disadvantage
of eliminating information from flux fluctuations on scales smaller
than $\Delta v$.  In particular, if the galaxy redshift error is
$\sigma_v=300 \kms$, then it makes sense to consider the average flux
over $\Delta v=300 \kms$ interval because the spectral points over
this interval are similarly likely to be at the true galaxy
redshift. We have estimated by eye from Figure 10 of
\cite{adelberger03} that the flux decrement of the same three galaxies
when averaged over $\Delta v=300 \kms$ around the galaxy redshift is
$0.29$. Figure 16 shows that the probability of this flux is only
$\sim 2\%$ indicating that this larger-scale feature is also not
easily explained by our simulated spectra.

In agreement with \cite{kollmeier03a}, we find that the small number
of galaxies in the observed sample do not {\it easily} explain the
discrepancy between the simulation predictions and the
\cite{adelberger03} data, even if we consider the dominant
contributions of galactic winds and redshift errors.  Fortunately, the
size of these kinds of data sets is bound to increase over the next
few years, and comparison of larger data sets to predictions of the
type presented here should definitively show whether there is or is
not a contradiction between the models and the observations.

\section{Discussion and Conclusions}

  We have analyzed the \lya forest in the neighborhood of galaxies
using cosmological simulations of structure formation that include
galactic winds. We find that winds can greatly affect the temperature
structure of intergalactic gas that is relatively close to galaxies,
but that the density structure is less affected. Changes in the
structure of the optical depth giving rise to the \lya forest are
smaller still than in the density field because of the tendency of
peculiar velocities to smear out the changes in regions affected by
winds. These results are broadly in accordance with previous
studies that used SPH simulations with a variety of prescriptions for
modeling the effects of galactic winds \citep{theuns02, bruscoli03},
as well as with more simplified models \citep{croft02, kollmeier03a,
kollmeier03b, desjacques04}.

  Analysis of simulations with realistic wind prescriptions is
particularly interesting in light of recent observations of the
galaxy-\lya forest connection at $z=3$ \citep{adelberger03} and the
apparent detection of a ``galaxy proximity effect''.  While more data
are clearly needed in order to ensure that the result is not due to
the small number of galaxy-QSO pairs observed so far, we find that our
simulations are not successful in reproducing the magnitude of the
proximity effect observed, except as a rare statistical fluctuation.
Intrinsic galaxy velocity dispersions, caused by the motion of the
galaxies within the gaseous halos around them, may increase the
effective galaxy redshift error. This effect is highly uncertain from
a theoretical perspective, because it depends on the type of halos in
which LBGs form (in particular, whether LBGs tend to be massive
galaxies in halo centers, or faster-moving satellite galaxies), but
may make the current observations more likely.

\begin{figure*}
\centerline{
\epsfxsize=7.5truein
\epsfbox[29 150 569 694]{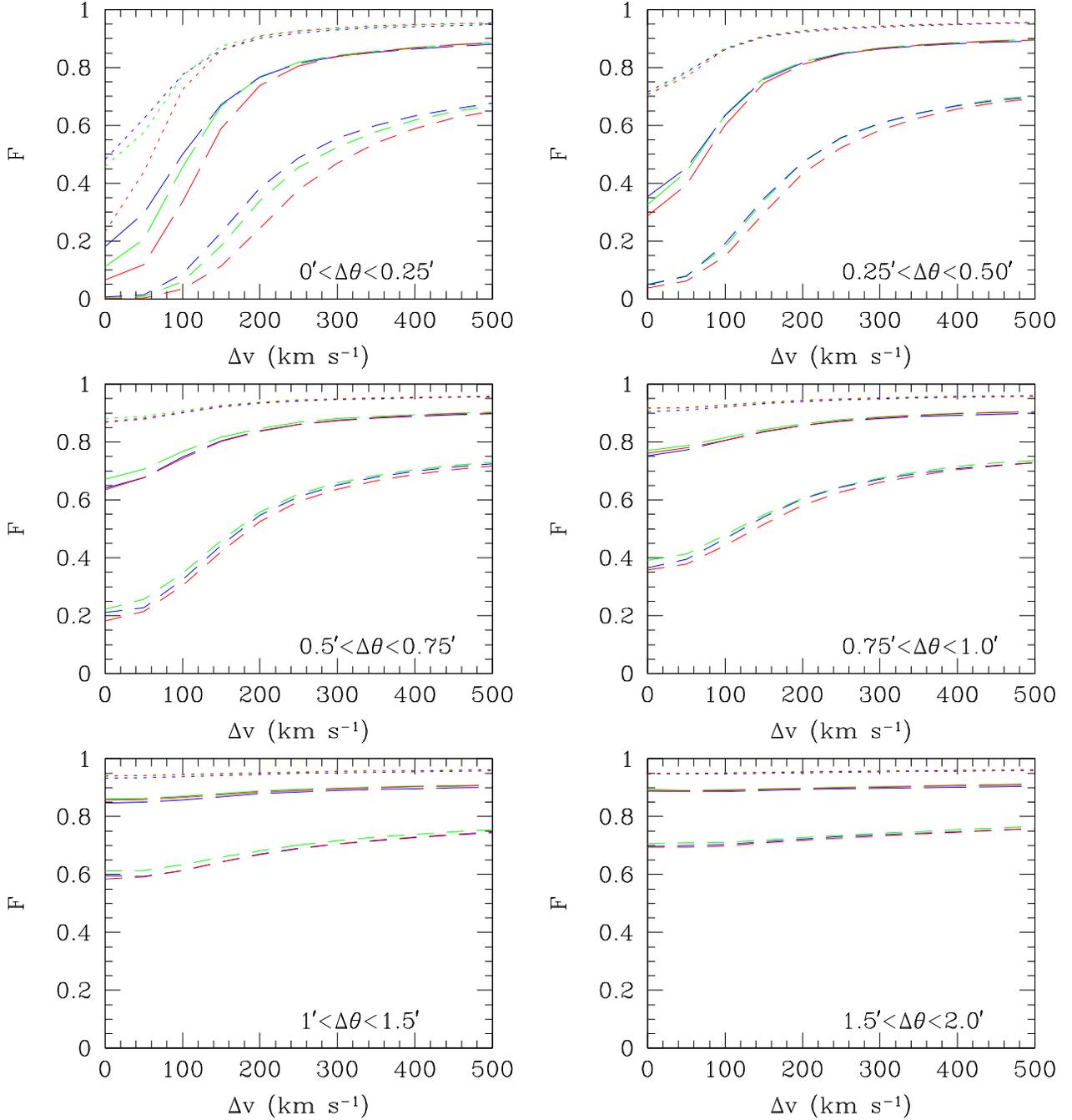}
}
\caption{Flux distribution as a function of galactic wind strength.
The $10\%$ (dotted),$25\%$ (long dashed), and $50\%$ (short dashed)
values of the flux as a function of galaxy proximity in pixels in
annuli of $\Delta\theta$ and averaged over a velocity interval
$\pm\Delta v$ of the 100 most massive galaxies in the L11-Low (red),
L11-Med (green), and L11-High (blue) simulations corresponding to
$\epsilon_{SN}=3\times 10^{-7}, 3\times 10^{-6}$ and $1.5\times
10^{-5}$ respectively. The likelihood of observing high transmitted
flux close to galaxies is greater in the High simulation indicating
that feedback has an effect on some pixels.}

\label{fig:wmappflux}
\end{figure*}

\begin{figure*}
\centerline{
\epsfxsize=7.5truein
\epsfbox[29 150 569 694]{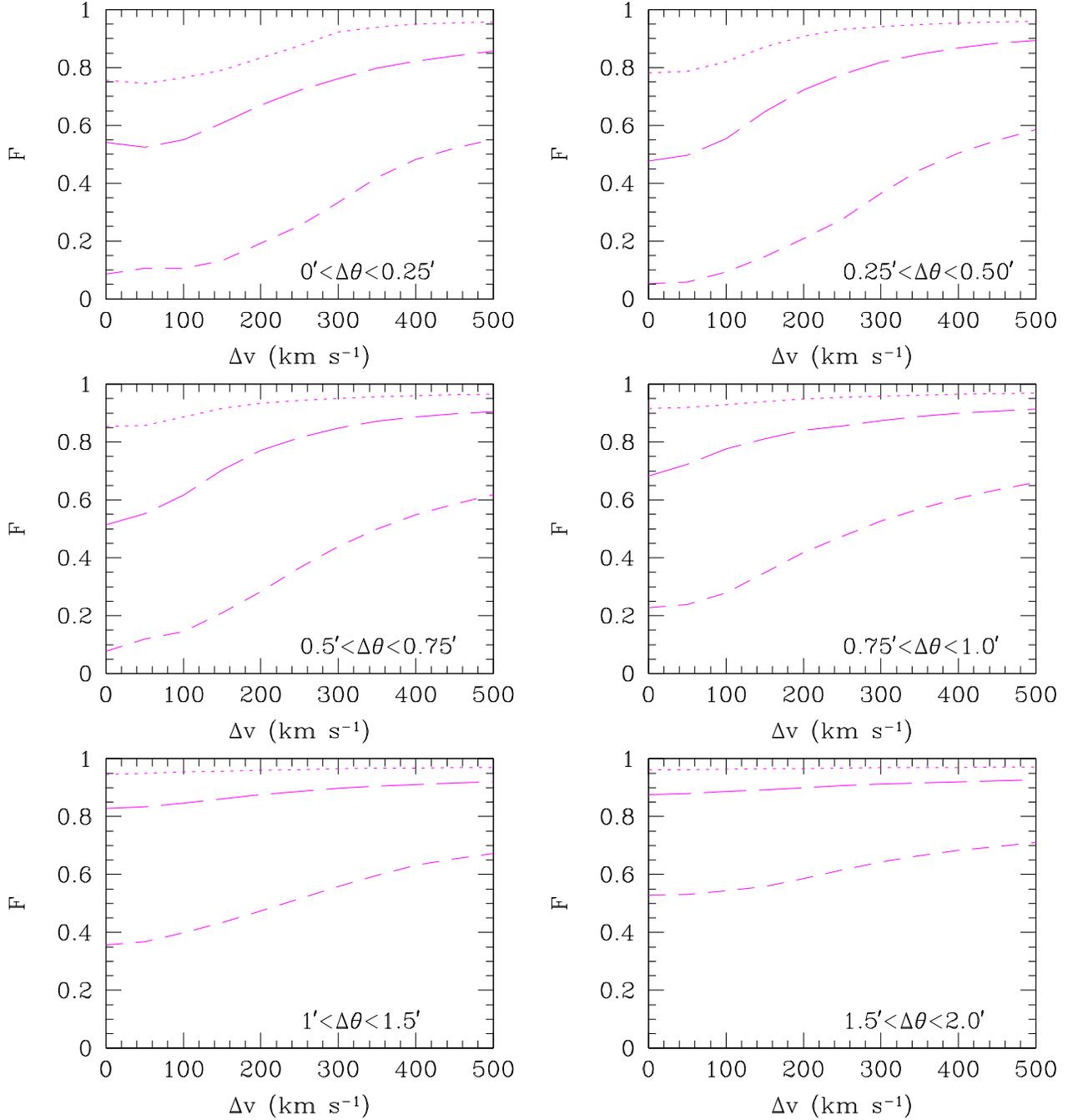}
}
\caption{Flux distribution for the L25 simulation (magenta curves).
The $10\%$ (dotted), median (short dashed), and $25\%$ (long dashed)
values of the flux as a function of galaxy proximity in pixels in
annuli of $\Delta\theta$ and averaged over $\pm\Delta v$ of the 100
most luminous galaxies in the simulation.  The box size and galaxy
number density affect these curves strongly. }

\label{fig:pfluxl25}
\end{figure*}

\begin{figure*}
\centerline{
\epsfxsize=4.5truein
\epsfbox[92 490 320 682]{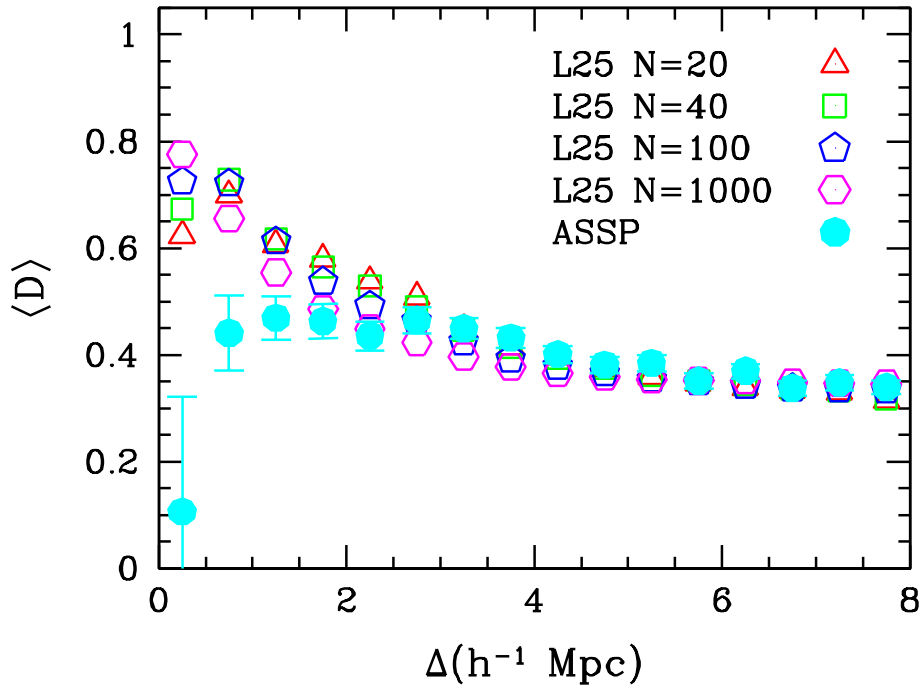}
}
\caption{Comparison of predictions from L25 with recent observations by
Adelberger et al. 2003.  Shown are the results for the mean flux
decrement $\langle D \rangle$ as a function of redshift-space
separation, $\Delta$ and space density for $N=20$(red triangles),$N=40$ (green squares), $N=100$ (blue pentagons) and $N=1000$ (magenta hexagons) simulations
compared with the Adelberger et al. 2003 observations (filled points).  The detailed shapes of these curves seems dependent on galaxy number density, although this may be due to the poorer statistics with the smaller galaxy samples.
}
\label{fig:adelcomp}
\end{figure*}

\begin{figure*}
\centerline{
\epsfxsize=6.5truein
\epsfbox[21 147 583 719]{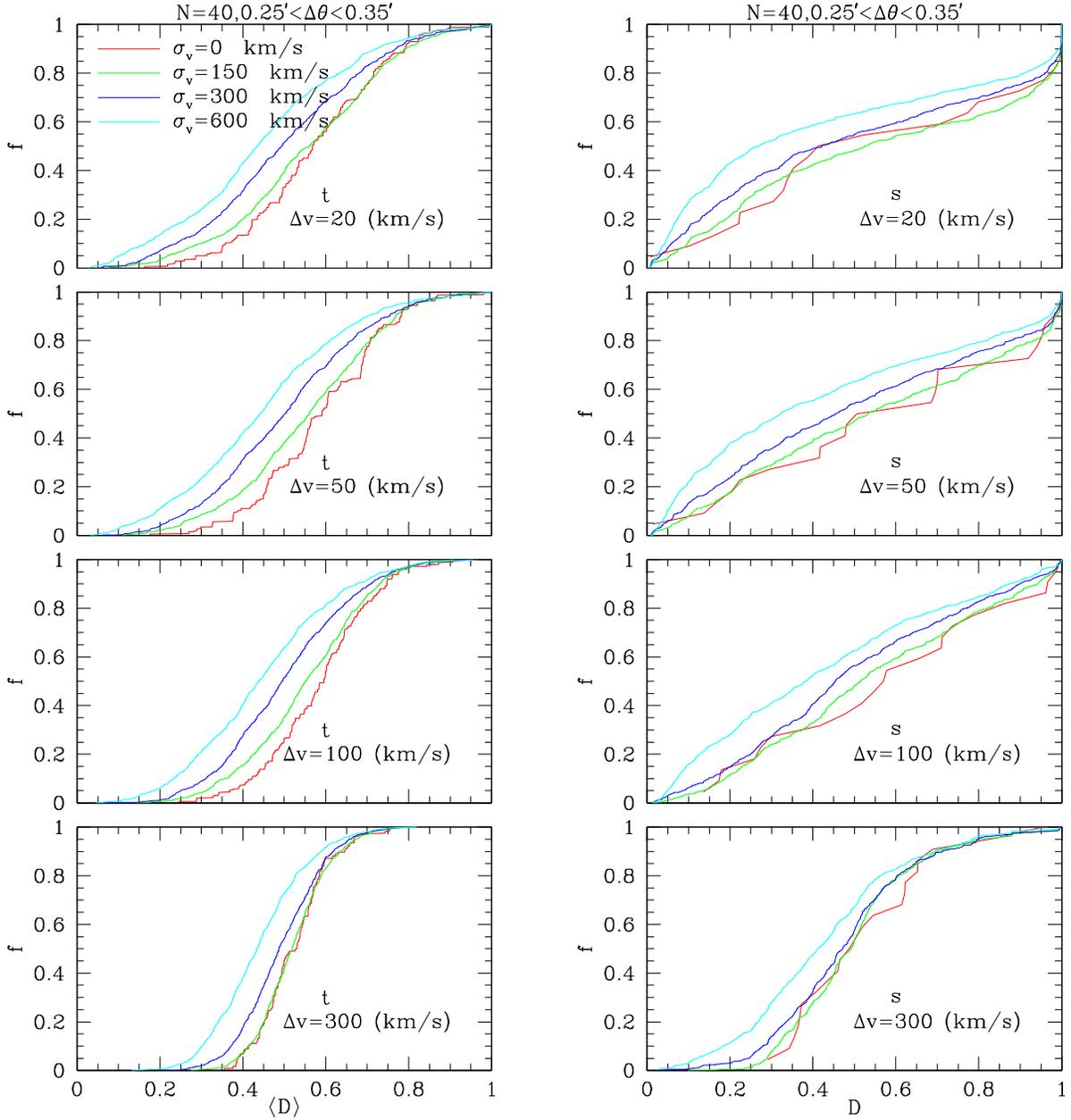}
}
\caption{Cumulative distribution of the average flux decrement near
LBGs as functions of redshift error and spectral resolution. Right
hand panels show the decrement, D, for individual galaxies within the
simulation and left hand panels show the average decrement, $\langle
D\rangle$ for galaxy triplets. Upper, middle, and lower panels show
the averages for velocity intervals $\Delta v=\pm20, 50, 100,300\kms$
respectively.  Curves show the cumulative distribution as a function
of galaxy redshift errors. Errors in the galaxy redshifts are drawn
from a Gaussian distribution of zero mean and dispersion, $\sigma$.
Red, green, blue and cyan curves show the cases for $\sigma_v = 0,
150, 300,$ and $600\kms$ respectively.  The galaxy sample is chosen
according to a luminosity threshold such that the observed number
density of LBGs is matched. The angular separation between the
galaxies and sightlines is chosen to correspond to the three closest
galaxy-QSO separations in \citet{adelberger03}.  }
\label{fig:fluke}
\end{figure*}

 The \lya flux decrement is generally predicted to increase near
galaxies owing to the high gas density in regions that have
gravitationally collapsed. Winds have the effect of diminishing the
magnitude of this increase. This effect of winds, although small, is
in principle detectable, in particular by complementing \lya
measurements with metal-line observations that could be sensitive to
the gas temperature, although we have not investigated the effect on
metal-lines in this paper. To detect the effect of winds in reducing
the \lya absorption near galaxies, accurate theoretical predictions
are required for the expected \lya flux decrement distribution as a
function of velocity and angular separation from a galaxy, for
different wind strengths. We have presented these predictions here,
although with the caveat that the predictions may depend on the
uncertain correspondence of observed Lyman break galaxies and the
galaxies in the simulations we analyze.  A large number of galaxies
with high signal-to-noise quasar spectra (similar to those used in the
\citealt{adelberger03} study) from nearby sightlines are essential to
search for the effects of winds. As shown in our Figures
\ref{fig:wmappflux} and \ref{fig:pfluxl25}, large redshift errors and
averaging the flux over large intervals in velocity will seriously
affect the measurement of the effect of winds on the galaxy-\lya
absorption correlation, since the signature of winds is strongest
close to the galaxies' systemic velocities.  Accurate determination of
the clustering properties of LBGs (or other high-redshift galaxy
samples in which the selection function is well understood) in
redshift space should allow us to correct for the effect of intrinsic
galaxy redshift errors on the observed galaxy-\lya correlation for
future comparisons.  With precise redshifts and high-resolution
spectra, the effect of winds at small separations is detectable.

  Ongoing observations will provide a large increase in the sample of
galaxies used to determine the mean \lya transmitted flux near LBGs
(Adelberger \etal, 2005). These observations will result in a dramatic
improvement in our assessment of the effects of winds on the \lya
forest observed near galaxies.

\acknowledgments{We are grateful to Kurt Adelberger, Andy Gould, Pat
McDonald, Alice Shapley, Chuck Steidel, and David Weinberg for stimulating
discussions and comments on this work.  This work was supported in
part by grants NSF-0098515, AST-0098515, AST-0201266, AST-0206299,
AST-0407176 and NAG5-13381.}

\begin{deluxetable}{lccccccccc}
\tablecolumns{10}
%\tablewidth{10pc}
\scriptsize
\tablecaption{Simulation Parameters}
\tablehead{ \colhead{Name} & \colhead{$L_{\rm box}$ $(\hmpc)$ \tablenotemark{a}} & \colhead{$N_{cell}$} & \colhead{$\epsilon_{\rm SN}$ } & \colhead{$\Omega_m$} & \colhead{$\Omega_{\Lambda}$ } & \colhead{$\Omega_b$} & \colhead{$h$} &\colhead{$\sigma_8$} & \colhead{$n$}}
\startdata

L11-High  & $11$ & $432^{3}$ & $1.5 \times 10^{-5}$  & $0.29$   & $0.71$ & $0.047$ & 0$.70$ & $0.85$ & $1.0$
\\
L11-Medium  & $11$ & $432^{3}$ & $3 \times 10^{-6}$  & $0.29$   & $0.71$ & $0.047$ & 0$.70$ & $0.85$ & $1.0$
\\
L11-Low  & $11$ & $432^{3}$ & $3 \times 10^{-7}$  & $0.29$   & $0.71$ & $0.047$ & 0$.70$& $0.85$ & $1.0$
\\
L25 & $25$ & $768^{3}$ & $1 \times 10^{-5}$  & $0.30$   & $0.70$ & $0.035$ & 0$.67$ & $0.9$ & $1.0$
\enddata
\tablenotetext{a}{
Length units are comoving.
}
\label{tab:sims}
\end{deluxetable}

\

\end{document}